\documentclass[sigconf]{acmart}

\usepackage{booktabs} 

\usepackage{amsmath,amsthm,amsfonts,amssymb}
\usepackage{hyperref}
\usepackage{tikz}

\usepackage{xspace,url}
\usepackage{amsmath,amssymb,hyperref}

\usepackage{amssymb,amsmath,graphics,graphicx,latexsym,amssymb,url,amsthm}
\usepackage{balance}

\usepackage{times}

\usepackage{hyperref}

\usepackage{lipsum,algorithm,multicol,url,mdframed}

\usepackage{amssymb}
\usepackage{graphicx}
\usepackage{algpseudocode}

\usepackage{pifont}
\newcommand{\cmark}{\ding{51}}%
\newcommand{\xmark}{\ding{55}}%

\newcommand{\objective}{\ensuremath{\mathcal{O}}}
\newcommand{\program}{\ensuremath{\mathcal{P}}}
\newcommand{\cache}{\ensuremath{\mathcal{C}}}
\newcommand{\pred}{\ensuremath{\mathit{Pred}}}

\renewcommand{\textrightarrow}{$\rightarrow$}

\setcopyright{rightsretained}

\usepackage{amssymb}
\usepackage{graphicx}
\usepackage{algpseudocode}

\usepackage{pifont}

\usepackage{fp}
\usepackage{color,xcolor}
\usepackage{colortbl}

\definecolor{LightGreen}{rgb}{0.7, 1.0, 0.7}
\definecolor{LightRed}{rgb}{1.0, 0.7, 0.7}
\definecolor{LightBlue}{rgb}{0.7, 0.7, 1.0}
\definecolor{TableHeader}{rgb}{0.9, 0.9, 0.9}

\newtheorem{app}{Property}
\newtheorem{app1}{Definition}

\newmdenv[innerlinewidth=0.5pt, roundcorner=4pt,linecolor=gray,innerleftmargin=8pt,
innerrightmargin=8pt,innertopmargin=8pt,innerbottommargin=8pt]{note}

\newcommand{\CC}{\textsc{Cache\-Fix}\xspace}

\renewcommand\footnotetextcopyrightpermission[1]{} 
\begin{document}


\title[CACHEFIX]{Symbolic Verification of Cache Side-channel Freedom}

\author{Sudipta Chattopadhyay}
\affiliation{%
\institution{Singapore University of Technology and Design}}

\author{Abhik Roychoudhury}
\affiliation{%
\institution{National University of Singapore}}


\begin{abstract}
Cache timing attacks allow third-party observers to retrieve sensitive
information from program executions. But, is it possible to automatically
check the vulnerability of a program against cache timing attacks and
then, automatically shield program executions against these attacks?
For a given program, a cache configuration and an attack model, our
\CC framework either verifies the cache side-channel freedom of the
program or synthesizes a series of patches to ensure cache side-channel
freedom during program execution. At the core of our framework is a
novel symbolic verification technique based on automated abstraction
refinement of cache semantics.
The power of such a framework is to allow symbolic reasoning over
counterexample traces and to combine it with runtime monitoring for
eliminating cache side channels during program execution. Our evaluation
with routines from \texttt{OpenSSL}, \texttt{libfixedtimefixedpoint},
\texttt{GDK} and \texttt{FourQlib} libraries reveals that our \CC approach
(dis)proves cache side-channel freedom within an average of 75 seconds.
Besides, in all except one case, \CC synthesizes {\em all} patches
within 20 minutes to ensure cache side-channel freedom of the respective
routines during execution.
\end{abstract}

%
%




\clubpenalty=10000
\widowpenalty=10000 

\maketitle

\section{Introduction}
\label{sec:introduction}
Cache timing attacks~\cite{Kocher2018spectre,Lipp2018meltdown} are among 
the most critical {\em side-channel attacks}~\cite{side-channel-survey-paper} 
that retrieve sensitive information from program executions. Recent cache 
attacks~\cite{armageddon} further show that cache side-channel attacks are 
practical even in commodity embedded processors, such as in ARM-based 
embedded platforms~\cite{armageddon}. 
The basic idea of a cache timing attack is to observe the timing of cache 
hits and misses for a program execution. Subsequently, the attacker use 
such timing to guess the sensitive input via which the respective program 
was activated. 

Given the practical relevance, it is crucial to verify whether a given program 
({\em e.g.} an encryption routine) satisfies {\em cache side-channel freedom}, 
meaning the program {\em is not vulnerable} to cache timing attacks. However, 
verification of such a property is challenging for several reasons. Firstly, 
the verification of cache side-channel freedom requires a systematic integration 
of cache semantics within the program semantics. This, in turn, is based on 
the derivation of a suitable abstraction of cache semantics. Our proposed 
\CC approach automatically builds such an abstraction and systematically refines 
it until a proof of cache side-channel freedom is obtained or a real 
({\em i.e.} non-spurious) counterexample is produced. Secondly, proving cache 
side-channel freedom of a program requires reasoning over multiple execution 
traces. To this end, we propose a symbolic verification technique 
within our \CC framework. Concretely, we capture the cache behaviour of a 
program via symbolic constraints over program inputs. Then, we leverage  
recent advances on {\em satisfiability modulo theory} (SMT) and constraint solving 
to (dis)prove the cache side-channel freedom of a program.

An appealing feature of our \CC approach is to employ symbolic reasoning 
over the real counterexample traces. To this end, we systematically explore 
real counterexample traces and apply such symbolic reasoning to synthesize 
patches. Each synthesized patch captures a symbolic condition $\nu$ on input 
variables and a sequence of actions that needs to be applied when the 
program is processed with inputs satisfying $\nu$. The application of a patch 
is guaranteed to reduce the channel capacity of the program under inspection. 
Moreover, if our checker terminates, then our \CC approach guarantees to 
synthesize all patches that {\em completely shields the program} against 
cache timing attacks~\cite{timing-attack-paper,trace-attack-paper}. 
Intuitively, our \CC approach can start with a program $\program$ vulnerable 
to cache timing attack. Then, it leverages a systematic combination of symbolic 
verification and runtime monitoring to execute $\program$ with cache 
side-channel freedom.

It is the precision and the novel mechanism implemented within \CC that 
set us apart from the state of the art. Existing works on analyzing cache 
side channels~\cite{cacheaudit,post17-paper,cav12-paper} are incapable 
to automatically build and refine abstractions for cache semantics. 
Besides, these works are not directly applicable when the underlying program 
{\em does not} satisfy cache side-channel freedom. Given an arbitrary 
program, our \CC approach generates proofs of its cache side-channel 
freedom or generates input(s) that manifest the violation of cache 
side-channel freedom. Moreover, our symbolic reasoning framework provides 
capabilities to systematically synthesize patches and completely eliminate 
cache side channels during program execution. 

We organize the remainder of the paper as follows. After providing an 
overview of \CC (\autoref{sec:motivation}), we make the following 
contributions: 

\begin{enumerate}
\item We present \CC, a novel symbolic verification framework to check 
the cache side-channel freedom of an arbitrary program. To the best of 
our knowledge, this is the first application of automated abstraction 
refinement and symbolic verification to check the cache behaviour of 
a program. 

\item We instantiate our \CC approach with direct-mapped caches, as well 
as with set-associative caches with {\em least recently used} (LRU) and 
{\em first-in-first-out} (FIFO) policy (\autoref{sec:cache-semantics}). 
In \autoref{sec:side-channel-property}, we show the generalization 
of our \CC approach over timing-based attacks~\cite{timing-attack-paper} 
and trace-based attacks~\cite{trace-attack-paper}. 

\item We discuss a systematic exploration of counterexamples to synthesize 
patches and to shield program executions against cache timing attacks 
(\autoref{sec:monitor}). We provide theoretical guarantees that such patch 
synthesis converges towards completely eliminating cache side channels 
during execution. 

\item We provide an implementation of \CC and evaluate it with 25 
routines from \texttt{OpenSSL}, \texttt{GDK}, \texttt{FourQlib} and 
\texttt{libfixedtimefixedpoint} libraries. Our evaluation reveals 
that \CC can establish proof or generate non-spurious counterexamples 
within 75 seconds on average. Besides, in most cases, \CC generated all 
patches within 20 minutes to ensure cache side-channel freedom during 
execution. Our implementation and all experimental data are publicly 
available. 

\end{enumerate}
 
\begin{figure*}[!htb]
\begin{center}
\begin{tabular}{cccc}
\rotatebox{0}{
\includegraphics[scale = 0.4]{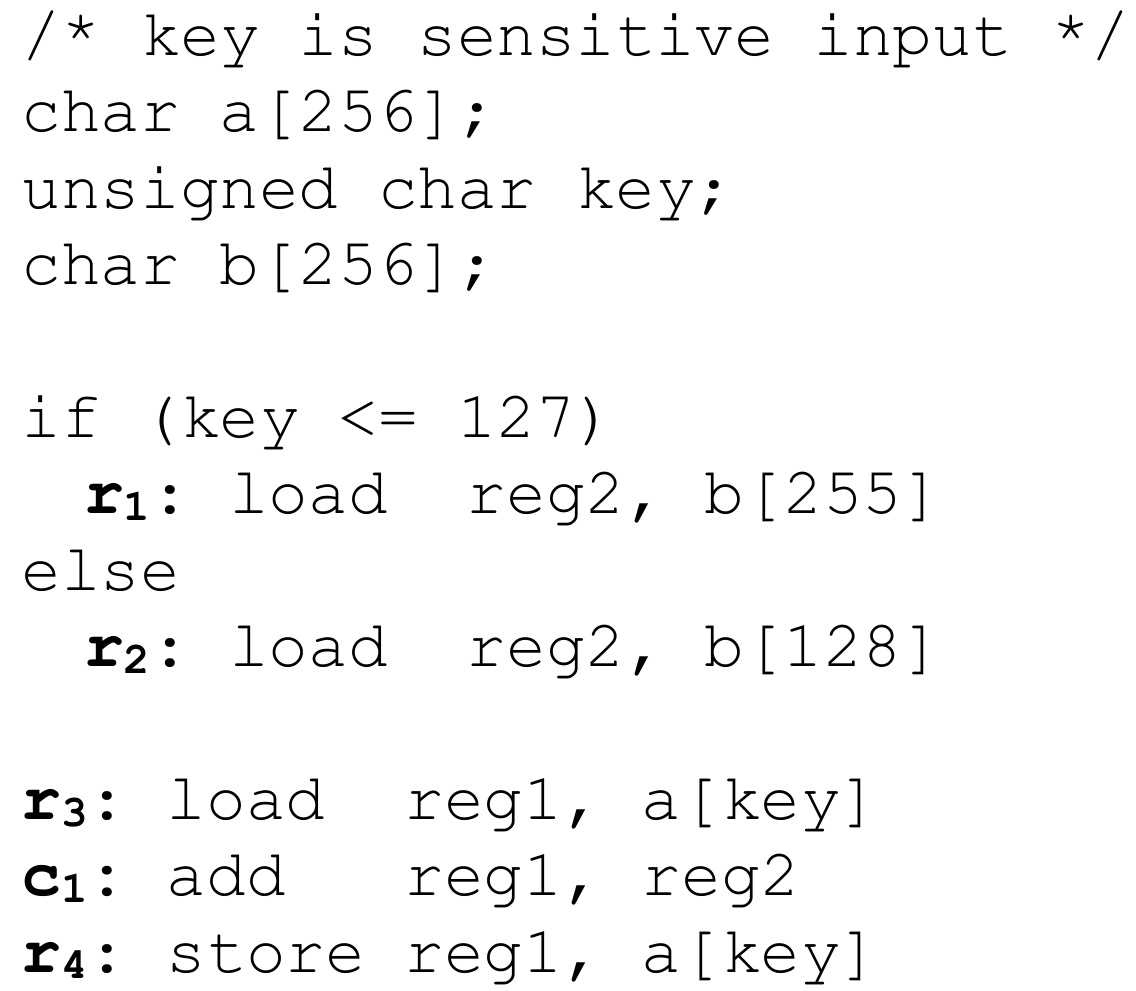}} & 
\rotatebox{0}{
\includegraphics[scale = 0.4]{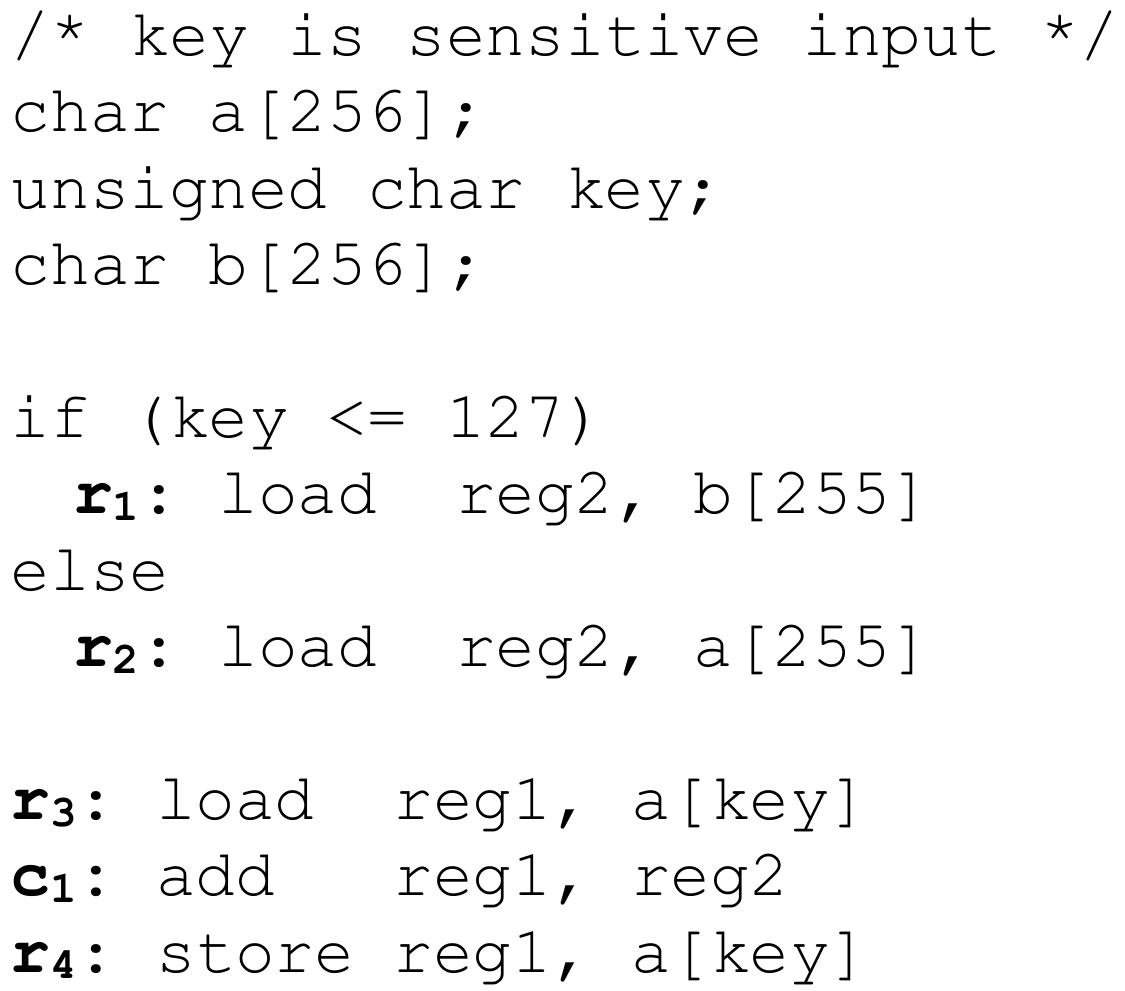}} & 
\rotatebox{0}{
\includegraphics[scale = 0.4]{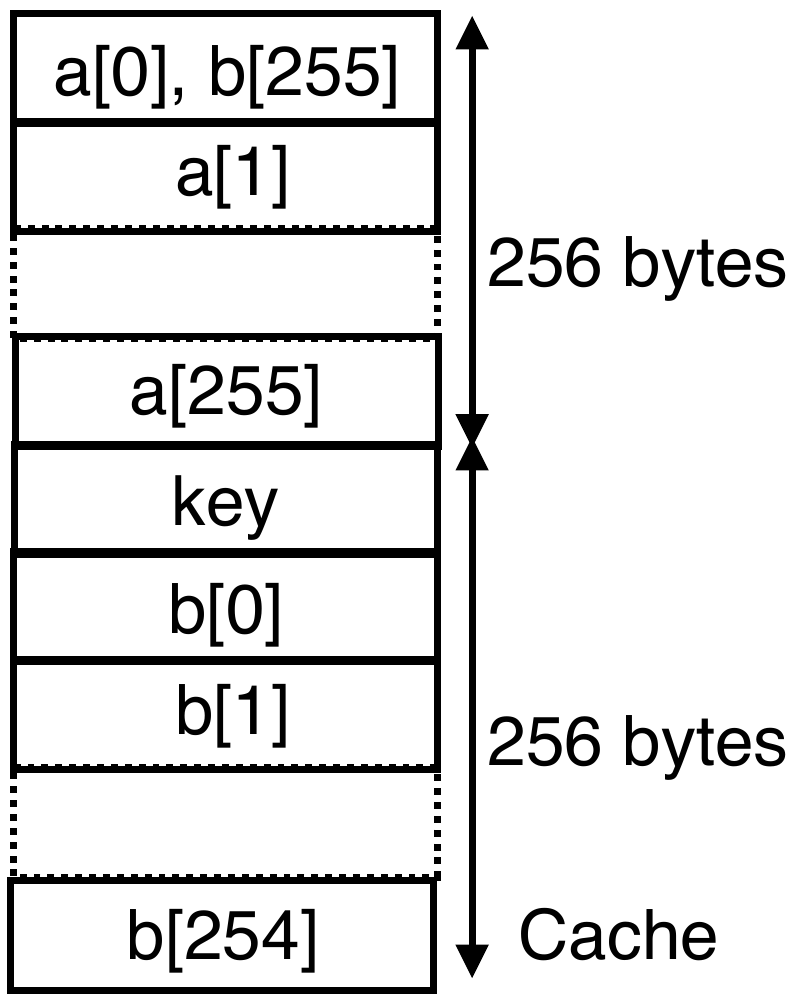}} & 
\rotatebox{0}{
\includegraphics[scale = 0.4]{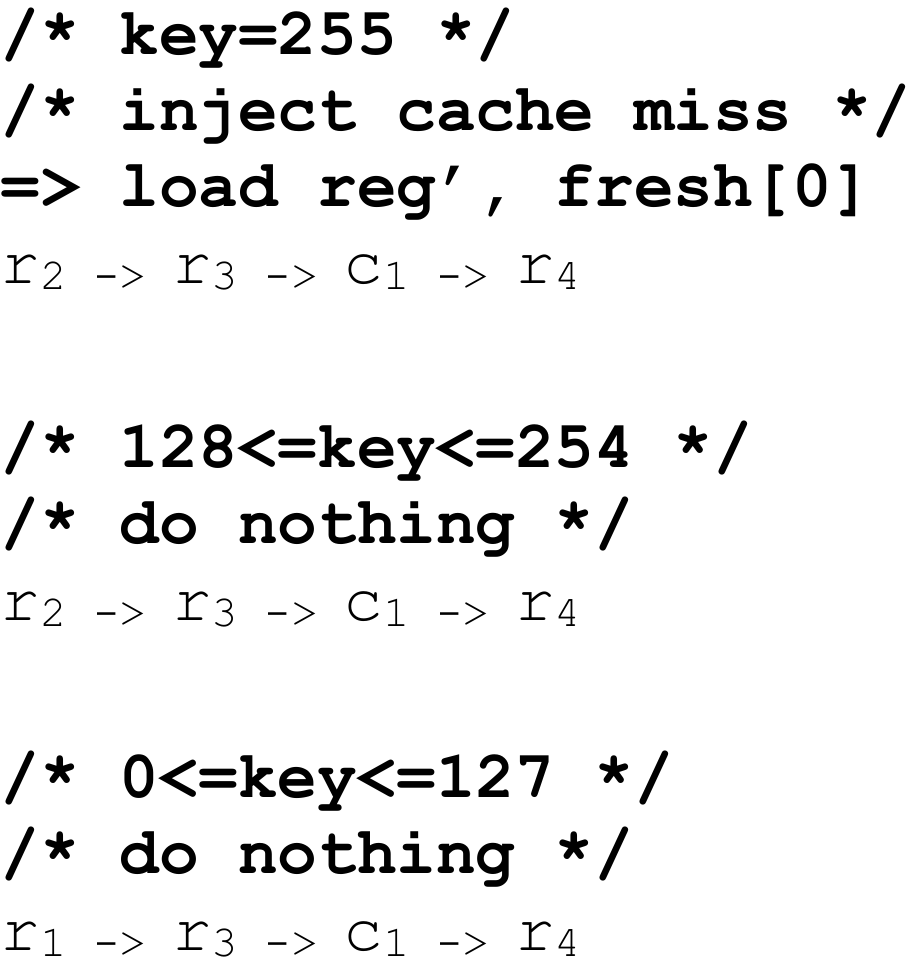}}
\\
\textbf{(a)} & \textbf{(b)} & \textbf{(c)} & \textbf{(d)}\\
\end{tabular}
\end{center}
\vspace*{-0.2in}
\caption{
 A code fragment (a) satisfying cache side-channel freedom, 
(b) violating cache side-channel freedom.
(c) Mapping of variables into the cache.
(d) Runtime actions and the execution order for the program in~\autoref{fig:example}(b) 
to ensure cache side-channel freedom.
}
\label{fig:example}
\vspace*{-0.1in}
\end{figure*}

\section{Overview}
\label{sec:motivation}



In this section, we demonstrate the general insight behind our approach through 
examples. We consider the simple code fragments in \autoref{fig:example}(a)-(b) 
where \texttt{key} is a sensitive input. In this example, 
we will assume a direct-mapped cache having a size of 512 bytes. For the sake of 
brevity, we also assume that $key$ is stored in a register and accessing $key$ 
does not involve the cache. The mapping of 
different program variables into the cache appears in \autoref{fig:example}(c). 
Finally, we assume the presence of an attacker who observes the number 
of cache misses in the victim program. 
For such an attacker, examples in \autoref{fig:example}(a)-(b) 
satisfy cache side-channel freedom if and only if the number of cache misses 
suffered is independent of \texttt{key}.




\paragraph*{\textbf{Why symbolic verification?}}
Cache side-channel freedom of a program critically depends on how it interacts with 
the cache. We make an observation that the program cache behaviour can be formulated 
via a well-defined set of predicates. To this end, let us assume $set(r_i)$ captures 
the cache set accessed by instruction $r_i$ and $tag(r_i)$ captures the accessed cache 
tag by the same instruction. Consider the instruction $r_3$ in \autoref{fig:example}(a). 
We introduce a symbolic variable $miss_3$, which we intend to set to one if $r_3$ 
suffers a cache miss and to set to zero otherwise. We observe that $miss_3$ depends 
on the following logical condition: 
\begin{equation}
\label{eq:example-gamma}
\begin{split}
\Gamma(r_3) \equiv \neg \left (0 \leq key \leq 127 \wedge \rho_{13}^{set} \wedge \neg \rho_{13}^{tag} \right ) 
\\
\wedge \neg \left ( key \ge 128 \wedge \rho_{23}^{set} \wedge \neg \rho_{23}^{tag} \right )
\end{split}
\end{equation}
where $\rho_{ji}^{tag} \equiv \left ( tag(r_j) \neq tag(r_i) \right )$ and 
$\rho_{ji}^{set} \equiv \left ( set(r_j) = set(r_i) \right )$. 
Intuitively, $\Gamma(r_3)$ checks whether both $r_1$ and $r_2$, if executed, load 
different memory blocks than the one accessed by $r_3$. Therefore, if $\Gamma(r_3)$ 
is evaluated to {\em true}, then $miss_3=1$ (i.e. $r_3$ suffers a cache miss) and 
$miss_3=0$ (i.e. $r_3$ is a cache hit), otherwise. Formally, we set the cache behaviour 
of $r_3$ as follows: 
\begin{equation}
\label{eq:example-miss}
\Gamma(r_3) \Leftrightarrow \left ( miss_3 = 1 \right );\ \ \neg \Gamma(r_3) \Leftrightarrow \left ( miss_3 = 0 \right )
\end{equation}
The style of encoding, as shown in \autoref{eq:example-miss}, facilitates the usage 
of state-of-the-art solvers for verifying cache side-channel freedom.

In general, we note that the cache behaviour of the program in \autoref{fig:example}(a), 
i.e., the cache behaviours of $r_1, \ldots ,r_4$; can be formulated accurately via the 
following set of predicates related to cache semantics: 
\begin{equation}
\label{eq:pred-set-example}
\pred_{cache} = \{\rho_{ji}^{set} \cup \rho_{ji}^{tag}\ |\ 1 \leq j < i \leq 4\}
\end{equation}

The size of $\pred_{cache}$ depends on the number of memory-related instructions. However, 
$\left | \pred_{cache} \right |$ does not vary with the cache size. 

\paragraph*{\textbf{Key insight in abstraction refinement}}
If the attacker observes the number of cache misses, then the cache side-channel 
freedom holds for the program in \autoref{fig:example}(a) when all feasible traces 
exhibit the same number of cache misses. Hence, such a property $\varphi$ can be 
formulated as the non-existence of two traces $tr_1$ and $tr_2$ as follows: 
\begin{equation}
\label{eq:property-example-trace}
\varphi \equiv \not \exists tr_1, \not \exists tr_2\ s.t.\ \left ( \sum_{i=1}^{4} miss_i^{(tr_1)} 
\ne \sum_{i=1}^{4} miss_i^{(tr_2)} \right )
\end{equation}
where $ miss_i^{(tr)}$ captures the valuation of $miss_i$ in trace $tr$. 

Our key insight is that to establish a proof of $\varphi$ (or its lack thereof), 
it is not necessary to accurately track the values of all predicates in 
$\pred_{cache}$ (cf. \autoref{eq:pred-set-example}). In other words, even if 
some predicates in $\pred_{cache}$ have {\em unknown} values, it might be possible 
to (dis)prove $\varphi$. This phenomenon occurs due to the inherent design principle 
of caches and we exploit this in our abstraction refinement process.

To realize our hypothesis, we first start with an initial set of predicates 
(possibly empty) whose values are accurately tracked during verification. In 
this example, let us assume that we start with an initial set of predicates 
$\pred_{init} = \{\rho_{13}^{set},  \rho_{13}^{tag}, \rho_{23}^{set}, \rho_{23}^{tag}\}$. 
The rest of the predicates in $\pred_{cache} \setminus \pred_{init}$ are set 
to {\em unknown} value. With this configuration at hand, \CC returns counterexample 
traces $tr_1$ and $tr_2$ (cf. \autoref{eq:property-example-trace}) to reflect that 
$\varphi$ does not hold for the program in \autoref{fig:example}(a). 
In particular, the following traces are returned: 
\begin{equation*}
\begin{split}
tr_1 \equiv \langle miss_1=miss_3=1, miss_2=miss_4=0 \rangle
\\
tr_2 \equiv \langle miss_1=0, miss_2=miss_3=miss_4=1 \rangle
\end{split}
\end{equation*}

Given $tr_1$ and $tr_2$, we check whether any of them are spurious. To this end, 
we reconstruct the logical condition (cf. \autoref{eq:example-miss}) that led 
to the specific valuations of $miss_i$ variables in a trace. For instance in 
trace $tr_2$, such a logical condition is captured via 
$\neg \Gamma(r_1) \wedge \bigwedge_{i \in [2,4]} \Gamma(r_i)$. It turns out that 
$\neg \Gamma(r_1) \wedge \bigwedge_{i \in [2,4]} \Gamma(r_i)$ 
is {\em unsatisfiable}, making $tr_2$ spurious. This happened due to the 
incompleteness in tracking the predicates $\pred_{cache}$.

To systematically augment the set of predicates and rerun our verification process, 
we extract the unsatisfiable core from 
$\neg \Gamma(r_1) \wedge \bigwedge_{i \in [2,4]} \Gamma(r_i)$. Specifically, we get the 
following unsatisfiable core: 
\begin{equation}
\label{eq:unsat-example}
\mathcal{U} \equiv \neg \rho_{34}^{set} \vee \rho_{34}^{tag} 
\end{equation}
Intuitively, with the initial abstraction $\pred_{init}$, our checker \CC 
failed to observe that $r_3$ and $r_4$ access the same memory block, hence, 
$\mathcal{U}$ is unsatisfiable. We then augment our initial set of predicates 
with the predicates in $\mathcal{U}$ and therefore, refining the abstraction 
as follows:
\begin{equation*}
\pred_{cur} = \{\rho_{13}^{set}, \rho_{13}^{tag}, \rho_{23}^{set}, \rho_{23}^{tag}, \rho_{34}^{set}, \rho_{34}^{tag}\}
\end{equation*}

\CC successfully verifies the cache side-channel freedom of the program in 
\autoref{fig:example}(a) with the set of predicates $\pred_{cur}$. We note 
that the predicates in $\pred_{cache} \setminus \pred_{cur} \ne \phi$. 
In particular, we still have {\em unknown} values assigned to the following 
set of predicates: 
\begin{equation*}
\pred_{unknown} = \{\rho_{12}^{set}, \rho_{12}^{tag}, \rho_{14}^{set}, \rho_{14}^{tag}, \rho_{24}^{set}, \rho_{24}^{tag}\}
\end{equation*}
Therefore, it was possible to verify $\varphi$ by tracking only half 
of the predicates in $\pred_{cache}$. Intuitively, $\rho_{12}^{set}$ 
and $\rho_{12}^{tag}$ were not needed to be tracked as $r_1$ and $r_2$ 
cannot appear in a single trace, as captured via the program 
semantics. In contrast, the rest of the predicates in $\pred_{unknown}$ 
were not required for the verification process, as neither $r_1$ nor 
$r_2$ influences the cache behaviour of $r_4$ -- it is influenced 
completely by $r_3$. 

\paragraph*{\textbf{Key insight in fixing}}
In general, the state-of-the-art in fixing cache side-channel is to 
revert to constant-time programming style~\cite{usenix16-paper}. 
Constant-time programming style imposes heavy burden on a programmer 
to follow certain programming patterns, such as to ensure the absence 
of input-dependent branches and input-dependent memory accesses. Yet, 
most programs do not exhibit constant-time behaviour. Besides, the 
example in \autoref{fig:example}(a) shows that an application can 
still have constant cache-timing, despite not following the 
constant-time programming style. Using our \CC approach, we observe 
that it is not necessary to always write constant-time programs. 
Instead, the executions of such programs can be manipulated to 
exhibit constant time behaviour. We accomplish this by leveraging 
our verification results. 

We consider the example in \autoref{fig:example}(b) and let us assume 
that we start with the initial abstraction 
$\pred_{init} = \{\rho_{13}^{set}, \rho_{13}^{tag}, \rho_{23}^{set}, \rho_{23}^{tag}\}$. 
\CC returns the following counterexample while verifying $\varphi$ 
(cf. \autoref{eq:property-example-trace}):
\begin{equation*}
\begin{split}
tr_1 \equiv \langle miss_1=miss_3=1, miss_2=miss_4=0 \rangle
\\
tr_2 \equiv \langle miss_2=1, miss_1=miss_3=miss_4=0 \rangle
\end{split}
\end{equation*}

If we reconstruct the logical condition that led to the specific valuations 
of $miss_1,\ldots,miss_4$ in $tr_1$ and $tr_2$, then we get the symbolic 
formulas 
$\Gamma(r_1) \wedge \neg \Gamma(r_2) \wedge \Gamma(r_3) \wedge \neg \Gamma(r_4)$ 
and 
$\neg \Gamma(r_1) \wedge  \Gamma(r_2) \wedge \neg \Gamma(r_3) \wedge \neg \Gamma(r_4)$, 
respectively. Both the formulas are satisfiable for the example in 
\autoref{fig:example}(b). Intuitively, this happens due to $r_2$, which 
loads the same memory block as accessed by $r_3$ only if $key=255$.

We observe that $tr_2$ will be equivalent to $tr_1$ if a cache miss is inserted 
in the beginning of $tr_2$. To this end, we need to know all inputs that lead 
to $tr_2$. Thanks to the symbolic nature of our analysis, we obtain the exact 
symbolic condition, i.e., 
$\neg \Gamma(r_1) \wedge  \Gamma(r_2) \wedge \neg \Gamma(r_3) \wedge \neg \Gamma(r_4)$ 
that manifests the trace $tr_2$. Therefore, if the program in \autoref{fig:example}(b) 
is executed with any input satisfying 
$\neg \Gamma(r_1) \wedge  \Gamma(r_2) \wedge \neg \Gamma(r_3) \wedge \neg \Gamma(r_4)$, 
then a cache miss is injected as shown in \autoref{fig:example}(d). This ensures 
the cache side-channel freedom during program execution, as all traces exhibit the same 
number of cache misses. 

\begin{figure}[t]
\centering
\includegraphics[scale= 0.28]{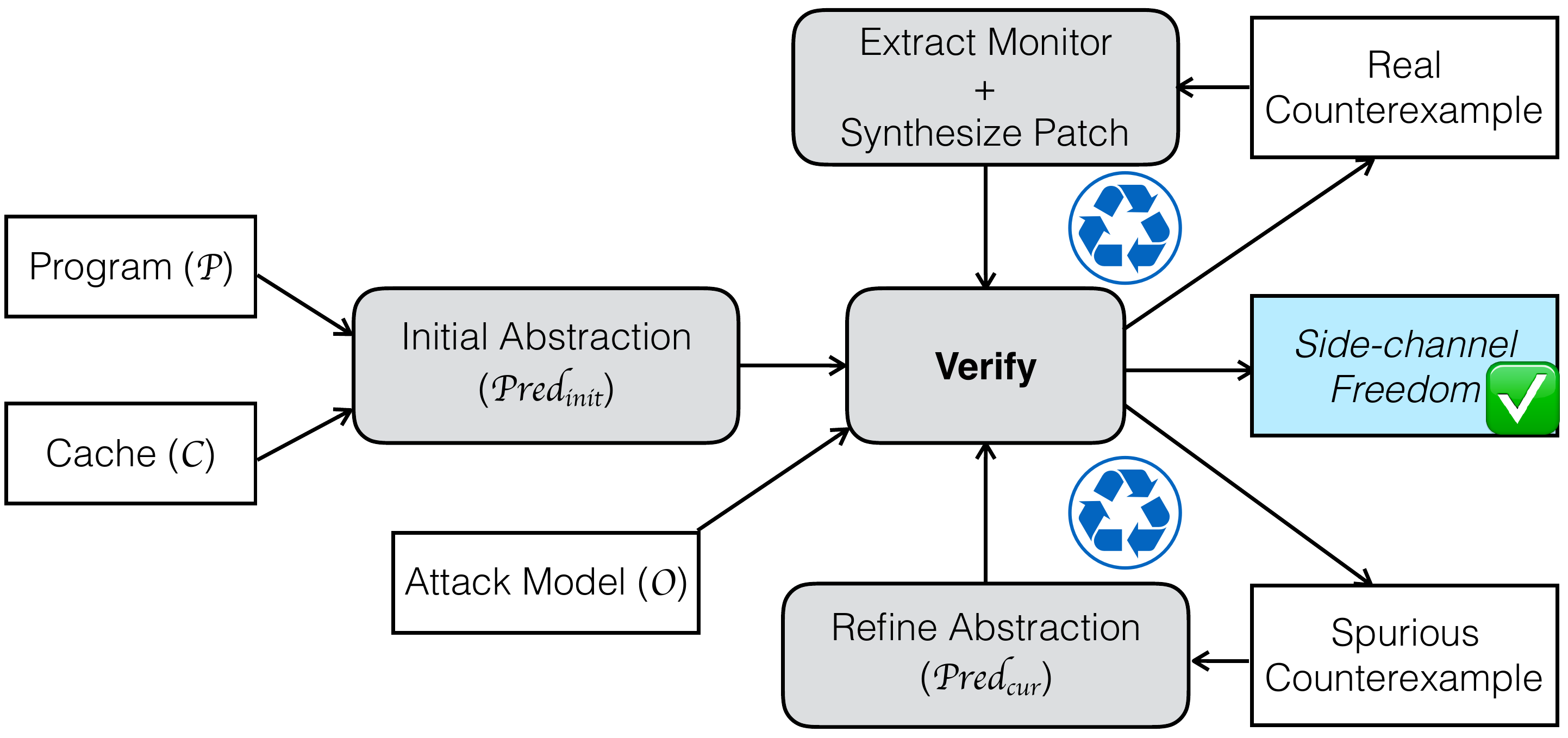}
\caption{\small Workflow of our symbolic verification and patching} 
\label{fig:workflow}
\end{figure}

Our proposed fixing mechanism is novel that it does not rely on any specific programming 
style. Moreover, as we generate the fixes by directly leveraging the verification results, 
we can provide strong security guarantees during program execution.

\paragraph*{\textbf{Overall workflow of \CC}}
\autoref{fig:workflow} outlines the overall workflow of \CC. The abstraction 
refinement process is guaranteed to converge towards the most precise 
abstraction of cache semantics to (dis)prove the cache side-channel freedom. 
Moreover, as observed in \autoref{fig:workflow}, our cache side channel fixing 
is guided by program verification output, enabling us to give cache side channel 
freedom guarantees about the fixed program.

\section{Threat and System Model}
\label{sec:workflow}

\paragraph*{\textbf{Threat Model}}
We assume that an attacker makes observations on the execution traces of victim 
program $\program$ and the implementation of $\program$ is known to the attacker. 
Besides, there does not exist any {\em error} in the observations made by the 
attacker. We also assume that an attacker can execute arbitrary user-level 
code on the processor that runs the victim program. This, in turn, allows the 
attacker to flush the cache ({\em e.g.} via accessing a large array) before 
the victim routine starts execution. We, however, do not assume that the attacker 
can access the address space of the victim program $\program$. We believe the 
aforementioned assumptions on the attacker are justified, as we aim to verify 
the cache side-channel freedom of programs against {\em strong attacker models}.  

We capture an execution trace via a sequence of hits ($h$) and misses ($m$).
Hence, formally we model an attacker as the mapping 
$\objective: \{h,m\}^{*} \rightarrow \mathbb{X}$, where $\mathbb{X}$ is a countable 
set. For ${tr}_1, {tr}_2 \in \{h,m\}^{*}$, an attacker can distinguish ${tr}_1$ from 
${tr}_2$ if and only if $\objective (tr_1) \ne \objective (tr_2)$. In this paper, we 
instantiate our checker for the following realistic attack models: 
\begin{itemize}
\item $\objective_{time}: \{h,m\}^{*} \rightarrow \mathbb{N}$. 
$\objective_{time}$ maps each execution trace to the number of cache 
misses suffered by the same. This attack model imitates cache timing 
attacks~\cite{timing-attack-paper}. 

\item $\objective_{trace}: \{h,m\}^{*} \rightarrow \{0,1\}^{*}$. 
$\objective_{trace}$ maps each execution trace to a bitvector 
($h$ is mapped to $0$ and $m$ is mapped to $1$). This attack 
model imitates trace-based attacks~\cite{trace-attack-paper}. 
\end{itemize}

\paragraph*{\textbf{Processor model}}
We assume an ARM-style processor with one or more cache levels. 
However, we consider timing attacks only due to first-level 
instruction or data caches~\cite{timing-attack-paper,trace-attack-paper}. 
We currently do not handle more advanced attacks on shared 
caches~\cite{gernot-cache-attack-paper}. First-level caches can either 
be partitioned (instruction vs. data) or unified. We assume that 
set-associative caches have either LRU or FIFO replacement policy. 
Other deterministic replacement policies can easily be integrated 
within \CC via additional symbolic constraints. Finally, our timing 
model only takes into account the effect of caches. Timing effects 
due to other micro-architectural features ({\em e.g.} pipeline 
and branch prediction) are currently not handled. For the sake of 
brevity, we discuss the timing effects due to memory-related instructions. 
It is straightforward to integrate the timing effects 
of computation instructions ({\em e.g.} \texttt{add}) into \CC.

\section{Abstraction Refinement}
\label{sec:abstraction}

\paragraph*{\textbf{Notations}} 
We represent cache via a triple 
$\langle 2^{\mathcal{S}}, 2^{\mathcal{B}}, \mathcal{A} \rangle$ where 
$2^{\mathcal{S}}$, $2^{\mathcal{B}}$ and $\mathcal{A}$ capture the 
number of cache sets, cache line size and cache associativity, respectively. 
We use $set(r_i)$ and $tag(r_i)$ to capture the cache set and cache tag, 
respectively, accessed by instruction $r_i$. 
%
%
%
Additionally, we introduce a symbolic variable $miss_i$ to capture whether 
$r_i$ was a miss ($miss_i = 1$) or a hit ($miss_i = 0$).
%
For instructions $r_i$ and $r_j$, we have $j < i$ if and only if $r_j$ 
was (symbolically) executed before $r_i$.

\subsection{Initial abstract domain}
We assume that a routine may start execution with any initial cache state, 
but it does not access memory blocks within the initial state during 
execution~\cite{cacheaudit}.
Hence, for a given instruction $r_i$, its cache behaviour might be affected by 
all instructions executing prior to $r_i$. Concretely, the cache 
behaviour of $r_i$ can be accurately predicted based on the set of logical 
predicates $\pred_{set}$ and $\pred_{tag}$ as follows:
\begin{equation}
\begin{split}
\pred_{set}^{i} = \{set(r_j)=set(r_i)\ |\ 1 \leq j < i\}
\\
\pred_{tag}^{i} = \{tag(r_j) \ne tag(r_i)\ |\ 1 \leq j < i\}
\end{split}
\end{equation}
Intuitively, $\pred_{set}^{i}$ captures the set of predicates checking 
whether any instruction prior to $r_i$ accesses the same cache set as 
$r_i$. Similarly, $\pred_{tag}^{i}$ checks whether any 
instruction prior to $r_i$ has a different cache tag than $tag(r_i)$. 
%
Based on this intuition, the following set of predicates 
are sufficient to predict the cache behaviours of $N$ 
memory-related instructions. 
\begin{equation}
\label{eq:all-predicates}
\pred_{set} = \bigcup_{i=1}^{N} \pred_{set}^{i};\ \ 
\pred_{tag} = \bigcup_{i=1}^{N} \pred_{tag}^{i}
\end{equation} 
For the sake of efficiency, however, we launch verification with 
a smaller set of predicates $\pred_{init} \subseteq \pred_{tag} \cup \pred_{set}$ 
as follows: 
\begin{equation}
\label{eq:init-abstraction}
\begin{split}
\pred_{init} = \bigcup_{i=1}^{N} \left \{ p \ |\ 
p \in \pred_{tag} \cup \pred_{set}\ \wedge |\sigma(r_i)|=1\ \wedge \right.
\\
\left. \forall k \in [1,i).\ |\sigma(r_k)|=1 \wedge guard_k \Rightarrow true \right \}  
\end{split}
\end{equation} 
$\sigma(r_i)$ captures the set of memory blocks accessed by instruction $r_i$ 
and $guard_k$ captures the control condition under which $r_k$ is executed. 
In general, our \CC approach works even if $\pred_{init} = \phi$. However, to 
accelerate the convergence of \CC, we start with the predicates whose values 
can be statically determined (i.e. independent of inputs). Intuitively, we take 
this approach for two reasons: firstly, the set $\pred_{init}$ can be computed 
efficiently during symbolic execution. Secondly, as the predicates in 
$\pred_{init}$ have constant valuation, they reduce the size of the formula to 
be discharged to the SMT solver.
%
We note that $guard_k$ depends on the program semantics. The abstraction 
of program semantics is an orthogonal problem and for the sake of brevity, 
we skip its discussion here. 
%

\subsection{Abstract domain refinement}
\label{sec:cache-abstraction}
We use the mapping $\Gamma: \{r_1,r_2,\ldots,r_N\} \rightarrow \{true,false\}$ 
to capture the conditions under which $r_i$ was a cache hit (i.e. $miss_i=0$) 
or a cache miss (i.e. $miss_i = 1$). 
In particular, the following holds: 
\begin{equation}
\label{eq:gamma-i}
\Gamma(r_i) \Leftrightarrow \left ( miss_i = 1 \right );\ \neg \Gamma(r_i) \Leftrightarrow \left ( miss_i = 0 \right ) 
\end{equation}
$\Gamma(r_i)$ depends on predicates in $\pred_{set}^{i} \cup \pred_{tag}^{i}$ 
and the cache configuration. We show the formulation of $\Gamma(r_i)$ in 
\autoref{sec:cache-semantics}.  


\begin{algorithm}[!b!t]
\caption{Abstraction Refinement Algorithm}
\textbf{Input:} Program $\program$, cache configuration $\cache$, 
attack model $\objective$\\
\textbf{Output:} Successful verification or a concrete counterexample
\label{alg:verification}
{
\begin{algorithmic}[1]
  \State{\textsf{/* $\Psi$ is a formula representation of $\program$ */}}
  \State{\textsf{/* $\pred$ is cache-semantics-related predicates */}}
  \State{\textsf{/* $\Gamma$ determines cache behaviour of all instructions */}}
  \State{$(\Psi,\pred,\Gamma)$ := \textsc{ExecuteSymbolic($\program$, $\cache$)}}	
  \State{\textsf{/* Formulate initial abstraction (cf. \autoref{eq:init-abstraction}) */}}
  \State{$\pred_{cur}$:=$\pred_{init}$ := \textsc{GetInitialAbstraction($\pred$)}}
  \State{\textsf{/* Rewrite $\Psi$ with initial abstraction */}}
  \State{\textsc{Rewrite$\left ( \Psi, \pred_{init} \right )$}}
  \State{\textsf{/* Formulate cache side-channel freedom property */}}
  \State{$\varphi$ := \textsc{GetProperty$\left ( \objective \right )$}}
  \State{\textsf{/* Invoke symbolic verification to check $\Psi \wedge \neg \varphi$ */}}
  \State{\label{ln:verify1}$(res, tr_1, tr_2)$ := \textsc{Verify}$\left ( \Psi , \varphi \right )$}
  \While{($res$=$false$) $\wedge$ ($tr_1$ or $tr_2$ is spurious)}
    \State{\small \textsf{/* Extract unsatisfiable core from $tr_1$ and/or $tr_2$ */}} 
  	\State{$\mathcal{U}$ := \textsc{UnsatCore$\left ( tr_1, tr_2, \Gamma \right )$}}
  	\State{\small \textsf{/* Refine abstractions and repeat verification */}} 
  	\State{$\pred_{cur}$ := \textsc{Refine}($\pred_{init}$, $\mathcal{U}$, $\pred$)}
  	\State{\textsc{Rewrite$\left ( \Psi, \pred_{cur} \right )$}}
  	\State{\label{ln:verify2}$(res, tr_1, tr_2)$ := \textsc{Verify}$\left ( \Psi , \varphi \right )$}
  	\State{$\pred_{init}$ := $\pred_{cur}$}
  \EndWhile
  \State{return $res$}
\end{algorithmic}}
\end{algorithm}

\floatname{algorithm}{Procedure}

\begin{algorithm}[t]
\caption{Symbolically Tracking Program and Cache States}
\label{alg:sym-exec}
{
\begin{algorithmic}[1]
  \State{\textsf{/* symbolically execute $\program$ with cache configuration $\cache$*/}}
  \Procedure{ExecuteSymbolic}{$\program$, $\cache$}
  	\State{$i$ := 1; $\Psi$ := $true$; $\pred_{set}$ := $\pred_{tag}$ := $\phi$}
  	\State{$r_i$ := \textsc{GetNextInstruction}($\program$)}
  	\While{$r_i \ne exit$}
  		\If{$r_i$ is memory-related instruction}
  			\State{\small \textsf{/* Collect predicates for cache semantics */}} 
  			\State{$\pred_{set}\ \cup= \pred_{set}^{i}$; $\pred_{tag}\ \cup= \pred_{tag}^{i}$}
  			\State{\small \textsf{/* $\Gamma(r_i)$ determines cache behaviour of $r_i$ */}} 
  			\State{Formulate $\Gamma(r_i)$ \textsf{/* see \autoref{sec:cache-semantics} */}}
  			\State{$\Gamma\ \cup= \{\Gamma(r_i)\}$}
  			\State{\small \textsf{/* Integrate cache semantics within $\Psi$ */}} 
  			\State{\label{ln:cache-start} $\Psi$ := \textsc{Convert}$\left ( \Psi, 
  			 \Gamma(r_i) \Leftrightarrow \left ( miss_i = 1 \right ) \right )$}
  			  			\State{\label{ln:cache-end} $\Psi$ := \textsc{Convert}$\left ( \Psi, 
  			 \neg \Gamma(r_i) \Leftrightarrow \left ( miss_i = 0 \right ) \right )$}
  		\EndIf
  		\State{\textsf{/* Integrate program semantics of $r_i$ within $\Psi$ */}} 
  		\State{\textsf{/* $\varphi(r_i)$ is a predicate capturing $r_i$ semantics */}} 
  		\State{\label{ln:program-semantics} $\Psi$ := \textsc{Convert}$\left ( \Psi, \varphi (r_i) \right )$}
  		\State{$i$ := $i+1$}
  		\State{$r_i$ := \textsc{GetNextInstruction}($\program$)}
  	\EndWhile
  	\State{return $(\Psi, \pred_{set} \cup \pred_{tag}, \Gamma)$}
  \EndProcedure
\end{algorithmic}}
\end{algorithm}

\floatname{algorithm}{Algorithm}

\paragraph*{\textbf{\textsc{ExecuteSymbolic}}}
Algorithm~\ref{alg:verification} captures the overall verification process based
on our systematic abstraction refinement. The symbolic verification engine computes 
a formula representation $\Psi$ of the program $\program$. 
This is accomplished via a symbolic execution on program $\program$ 
({\em cf.} procedure \textsc{ExecuteSymbolic}) and systematically 
translating the cache and program semantics of each instruction into a set of 
constraints ({\em cf.} procedure \textsc{Convert}). 
%

\paragraph*{\textbf{\textsc{Convert}}}
During the symbolic execution, a set of symbolic states, each capturing a unique 
execution path reaching an instruction $r_i$, is maintained. This set of symbolic 
states can be viewed as a disjunction 
$\Psi(r_i) \equiv \psi_1 \vee \psi_2 \vee \ldots \vee \psi_{j-1} \vee \psi_j$, 
where $\Psi(r_i) \Rightarrow \Psi$ and each $\psi_i$ symbolically captures a unique 
execution path leading to instruction $r_i$. At each instruction $r_i$, the 
procedure \textsc{Convert} translates $\Psi(r_i)$ in such a fashion that $\Psi(r_i)$ 
integrates both the cache semantics ({\em cf.} lines~\ref{ln:cache-start}-\ref{ln:cache-end}) 
and program semantics ({\em cf.} lines~\ref{ln:program-semantics}) of $r_i$. 
For instance, to integrate cache semantics of a memory-related instruction 
$r_i$, $\Psi(r_i)$ is converted to 
$\Psi(r_i) \wedge \left ( \Gamma(r_i) \Leftrightarrow \left ( miss_i=1 \right ) \right ) \wedge 
\left ( \neg \Gamma(r_i) \Leftrightarrow \left ( miss_i=0 \right ) \right )$.
Similarly, the program semantics of instruction $r_i$, as captured via $\varphi(r_i)$, 
is integrated within $\Psi(r_i)$ as $\Psi(r_i) \wedge \varphi(r_i)$. Translating 
the {\em program semantics} of each instruction to a set of constraints is a standard 
technique in any symbolic model checking~\cite{cimmati-paper}. Moreover, such a translation 
is typically carried out on a program in static single assignment (SSA) form and takes into 
account both data and control flow. Unlike classic symbolic analysis, however, we 
consider both the cache semantics and program semantics of an execution path, as explained 
in the preceding.

\paragraph*{\textbf{\textsc{GetInitialAbstraction}}}
We start our verification with an initial abstraction of cache semantics 
(cf. \autoref{eq:init-abstraction}). 
Such an initial abstraction contains a partial set of logical predicates 
$\pred_{init} \subseteq \pred_{set} \cup \pred_{tag}$. Based on $\pred_{init}$, 
we rewrite $\Psi$ via the procedure \textsc{Rewrite} as follows: We walk through 
$\Psi$ and look for occurrences of any predicate 
$p^{-} \in \left ( \pred_{set} \cup \pred_{tag} \right ) \setminus \pred_{init}$. 
For any $p^{-}$ discovered in $\Psi$, we replace $p^{-}$ with a fresh symbolic 
variable $V_{p^{-}}$. Intuitively, this means that during the verification process, 
we assume any truth value for the predicates in 
$\left ( \pred_{set} \cup \pred_{tag} \right ) \setminus \pred_{init}$. 
This, in turn, substantially reduces the size of the symbolic formula $\Psi$ 
and simplifies the verification process. 

\paragraph*{\textbf{\textsc{Verify} and \textsc{GetProperty}}}
The procedure \textsc{Verify} invokes the solver to check the cache side-channel 
freedom of $\program$ with respect to attack model $\objective$. The property 
$\varphi$, capturing the cache side-channel freedom, is computed via \textsc{GetProperty}. 
%
%
For example, in timing-based attacks, $\varphi$ is 
captured via the non-existence of any two traces $tr_1$ and $tr_2$ that have 
different number of cache misses (cf. \autoref{eq:property-example-trace}). 
In other words, if the following formula is satisfied with more than one 
valuations for $\sum_{i=1}^{N} miss_i$, then side-channel freedom is violated: 
\begin{equation}
\label{eq:timing-freedom-practice}
\Psi \wedge \left ( \left (  \sum_{i=1}^{N} miss_i \right ) \geq 0 \right ) 
\end{equation}
Here $N$ captures the total number of memory-related instructions encountered during 
the symbolic execution of $\program$. 

\paragraph*{\textbf{\textsc{Refine} and \textsc{Rewrite}}}
%
When our verification process fails, we check the feasibility of a 
counterexample trace $trace \in \{tr_1,tr_2\}$. Recall from \autoref{eq:gamma-i}
that $r_i$ is a cache miss if and only if $\Gamma(r_i)$ holds {\em true}. We 
leverage this relation to construct the following formula $\Gamma_{trace}$ 
for feasibility checking:
\begin{eqnarray}
\label{eq:counterexample}
\Gamma_{trace}  = \bigwedge_{i=1}^{N}
\begin{cases}
\Gamma(r_i),    	  \text{    if $miss_i^{(trace)} = 1$;}
\\
\neg \Gamma(r_i), \text{ if $miss_i^{(trace)} = 0$;}
\end{cases}
\end{eqnarray}
In \autoref{eq:counterexample}, $miss_i^{(trace)}$ captures the valuation of symbolic 
variable $miss_i$ in the counterexample $trace$. We note that $trace$ is not a spurious 
counterexample if and only if $\Gamma_{trace}$ is {\em satisfiable}, hence, highlighting 
the violation of cache side-channel freedom. 

If $\Gamma_{trace}$ is {\em unsatisfiable}, then our initial abstraction 
$\pred_{init}$ was insufficient to (dis)prove the cache side-channel freedom. 
In order to refine this abstraction, we extract the {\em unsatisfiable core} from 
the symbolic formula $\Gamma_{trace}$ via the procedure \textsc{UnsatCore}. Such 
an unsatisfiable core contains a set of CNF clauses 
$\in \bigcup_{k \in [1,N]} \Gamma(r_{k})$. 
We note each $\Gamma(r_{k})$ is a function of the set of predicates 
$\pred_{tag} \cup \pred_{set}$.
Finally, we refine the abstraction ({\em cf.} procedure \textsc{Refine} in 
Algorithm~\ref{alg:verification}) to $\pred_{cur}$ by including all 
predicates in the unsatisfiable core as follows:
%
%
\begin{equation}
\label{eq:refine}
\pred_{cur} := \pred_{init} \cup \{p^{+}\ |\ p^{+} \in \mathsf{UnsatCore}(\Gamma_{trace}) \setminus \pred_{init} \}
\end{equation}

With the refined abstraction $\pred_{cur}$, we rewrite the symbolic formula $\Psi$ 
({\em cf.} procedure \textsc{Rewrite}). 
In particular, we identify the placeholder symbolic variables 
for predicates in the set $\pred_{cur} \setminus \pred_{init}$. We rewrite $\Psi$ by 
replacing these placeholder symbolic variables with the respective 
predicates in the set $\pred_{cur} \setminus \pred_{init}$. It is worthwhile to note 
that the placeholder symbolic variables in 
$\left ( \pred_{tag} \cup \pred_{set} \right ) \setminus \pred_{cur}$ remain unchanged. 



\subsection{Modeling Cache Semantics}
\label{sec:cache-semantics}

For each memory-related instruction $r_i$, the formulation of $\Gamma(r_i)$ 
is critical to prove the cache side-channel freedom. The formulation of 
$\Gamma(r_i)$ depends on the configuration of caches. Due to space constraints, 
we will only discuss the symbolic model for direct-mapped caches (symbolic models 
for LRU and FIFO caches are provided in the appendix). 
%
To simplify the formulation, we will use the following abbreviations for the 
rest of the section: 
\begin{equation}
\rho_{ij}^{set} \equiv \left ( set(r_i) = set(r_j) \right ); \ \ 
\rho_{ij}^{tag} \equiv \left ( tag(r_i) \ne tag(r_j) \right )
\end{equation}
We also distinguish between the following variants of misses: 
\begin{enumerate}
\item \textbf{Cold misses:} Cold misses occur when a memory block is accessed 
for the first time. 

\item \textbf{Conflict misses:} All cache misses that are not cold misses are 
referred to as conflict misses. 
\end{enumerate}


\paragraph*{\textbf{Formulating conditions for cold misses}}
Cold cache misses occur when a memory block is accessed for {\em the first 
time during program execution}. 
In order to check whether $r_i$ suffers a cold miss, 
we check whether all instructions $r \in \{r_1, r_2, \ldots, r_{i-1}\}$ 
access different memory blocks than the memory block accessed by $r_i$. 
This is captured as follows:
\begin{equation}
\label{eq:cold}
\Theta_i^{cold} \equiv \bigwedge_{j \in [1,i)} 
\left ( \neg \rho_{ji}^{set} \vee \rho_{ji}^{tag} \vee \neg guard_j \right )
\end{equation}
Recall that $guard_j$ captures the control condition under which $r_j$ is executed. 
Hence, if $guard_j$ is evaluated false for a trace, then $r_j$ does not appear in 
the respective trace. If $\Theta_i^{cold}$ is satisfied, then 
$r_i$ inevitably suffers a cold cache miss. 

\paragraph*{\textbf{Formulating conditions for conflict cache misses}}
%
For direct-mapped caches, an instruction $r_i$ suffers a conflict miss due to 
an instruction $r_j$ if all of the following conditions are satisfied: 

$\mathbf{\phi_{ji}^{cnf,dir}}:$ If $r_j$ accesses the same cache set as $r_i$, however, 
$r_j$ accesses a different cache tag as compared to $r_i$. 
This is formally captured as follows: 
\begin{equation}
\label{eq:direct-map-dif}
\phi_{ji}^{cnf,dir} \equiv \rho_{ji}^{tag} \wedge \rho_{ji}^{set}
\end{equation}

$\mathbf{\phi_{ji}^{rel,dir}}:$ No instruction between $r_j$ and $r_i$ accesses the same 
memory block as $r_i$. For instance, consider the memory-block access sequence 
$(r_1:m_1) \rightarrow (r_2:m_2) \rightarrow (r_3:m_2)$, where both $m_1$ and $m_2$ 
are mapped to the same cache set and $r_{1 \ldots 3}$ captures the respective 
memory-related instructions. It is not possible for $r_1$ to inflict a conflict miss 
for $r_3$, as the memory block $m_2$ is reloaded by instruction $r_2$. 
$\phi_{ji}^{rel,dir}$ is formally captured as follows: 
\begin{equation}
\label{eq:reload}
\phi_{ji}^{rel,dir} \equiv \bigwedge_{j < k < i} \left ( \rho_{ki}^{tag} \vee \neg \rho_{ki}^{set} \vee \neg guard_k \right )
\end{equation} 
Intuitively, $\phi_{ji}^{rel,dir}$ captures that all instructions between $r_j$ and 
$r_i$ either access a different memory block than $r_i$ (hence, satisfying 
$ \rho_{ki}^{tag} \vee \neg \rho_{ki}^{set}$) or does not appear in the execution 
trace (hence, satisfying $\neg guard_k$).

Given the intuition mentioned in the preceding paragraphs, we conclude that $r_i$ 
suffers a conflict miss if both $\phi_{ji}^{cnf,dir}$ and $\phi_{ji}^{rel,dir}$ are 
satisfied for {\em any instruction executing prior to $r_i$}. This is captured in 
the symbolic condition $\Theta_{i}^{cnf,dir}$ as follows:
\begin{equation}
\label{eq:conflict-misses-dir}
\Theta_{i}^{cnf,dir} \equiv \bigvee_{j \in [1,i)} \left ( \phi_{ji}^{cnf,dir} \wedge \phi_{ji}^{rel,dir} \wedge guard_j \right ) 
\end{equation}

\paragraph*{\textbf{Computing $\Gamma(r_i)$}}
For direct-mapped caches, $r_i$ can be a cache miss if it is either a cold cache miss or 
a conflict miss. Hence, $\Gamma(r_i)$ is captured symbolically as follows: 
\begin{equation}
\label{eq:gamma-direct-mapped}
\boxed{
\Gamma(r_i) \equiv guard_i \wedge \left ( \Theta_{i}^{cold} \vee \Theta_{i}^{cnf,dir} \right )}
\end{equation}

\subsection{Property for cache side-channel freedom}
\label{sec:side-channel-property}
In this paper, we instantiate our checker for timing-based attacks~\cite{timing-attack-paper} 
and trace-based attacks~\cite{trace-attack-paper} as follows. 

\paragraph*{\textbf{Timing-based attacks}}
In timing-based attacks, an attacker aims to distinguish traces based on their timing. 
In our framework, we verify the following property to ensure cache side-channel freedom: 
\begin{equation}
\label{eq:timing-freedom-property}
\boxed{
\left | \Psi \wedge \left ( \left (  \sum_{i=1}^{N} miss_i \right ) \geq 0 \right ) \right |_{sol \left (\sum_{i=1}^{N} miss_i \right )} \leq 1}
\end{equation} 
$N$ captures the number of symbolically executed, memory-related instructions. 
$sol (\sum_{i=1}^{N} miss_i)$ captures the number of valuations of $\sum_{i=1}^{N} miss_i$.
Intuitively, 
\autoref{eq:timing-freedom-property} aims to check that the underlying program has exactly 
one cache behaviour, in terms of the total number of cache misses.

\paragraph*{\textbf{Trace-based attacks}}
In trace-based attacks, an attacker monitors the cache behaviour of each 
memory access. 
We define a partial function 
$\xi: \{r_1,\ldots,r_N\} \nrightarrow \{0,1\}$ as follows:
\begin{eqnarray}
\label{eq:xi}
\xi(r_i)  = 
\begin{cases}
1,    	  \text{    if $guard_i \wedge \left ( miss_i = 1 \right )$ holds;}
\\
0, \text{ if $guard_i \wedge \left ( miss_i = 0 \right )$ holds;}
\end{cases}
\end{eqnarray}
The following verification goal ensures side-channel freedom: 
%
{
\begin{equation}
\label{eq:trace-freedom-property}
\boxed{
\left | \Psi \wedge \left | dom(\xi) \right | \ge 0\ \wedge \left ( \parallel_{r_i \in dom(\xi)} \xi(r_i) \right ) 
\ge 0 \right |_{sol(X)} \leq 1}
\end{equation}
where $dom(\xi)$ captures the domain of $\xi$, $\parallel$ captures the ordered 
(with respect to the indexes of $r_i$) concatenation operation  and 
$X=\langle \left | dom(\xi) \right |, \parallel_{r_i \in dom(\xi)} \xi(r_i) \rangle$. 
Intuitively, we check whether there exists exactly one cache behaviour sequence.

\section{Runtime Monitoring}
\label{sec:monitor}

\CC produces the first real counterexample when it discovers two traces with 
different observations (w.r.t. attack model $\objective$). 
These traces are then analyzed to compute a set of runtime actions that are 
guaranteed to reduce the uncertainty to guess sensitive inputs. Overall, our 
runtime monitoring involves the following crucial steps: 
\begin{itemize}
\item We analyze a counterexample trace $tr$ and extract the symbolic 
condition for which the same trace would be generated, 
\item We systematically explore unique counterexamples with the objective 
to reduce the uncertainty to guess sensitive inputs,
\item We compute a set of runtime actions that need to be applied for improving 
the cache side-channel freedom.
\end{itemize} 
In the following, we discuss these three steps in more detail. 

\paragraph*{\textbf{Analyzing a counterexample trace}}
Given a real counterexample $trace$, we extract a symbolic condition 
that captures all the inputs for which the same counterexample $trace$ 
can be obtained. Thanks to the symbolic nature of our analysis, \CC 
already includes capabilities to extract these monitors as follows. 
\begin{eqnarray}
\label{eq:monitor}
\nu  \equiv \bigwedge_{r_i \in trace}
\begin{cases}
\Gamma(r_i),    	  \text{    if $miss_i^{(trace)} = 1$;}
\\
\neg \Gamma(r_i), \text{ if $miss_i^{(trace)} = 0$;}
\end{cases}
\end{eqnarray}
where 
$miss_i^{(trace)}$ is the valuation of symbolic variable $miss_i$ in $trace$. 
We note that $\nu \Rightarrow \neg \Gamma(r_j)$ for any $r_j$ 
that does not appear in $trace$ (i.e. $r_j \notin trace$). Hence, to formulate 
$\nu$, it was sufficient to consider only the instructions that appear in $trace$. 

%
%
Once we extract a monitor $\nu$ from counterexample $trace$, the symbolic system 
$\Psi$ is refined to $\Psi \wedge \neg \nu$. This is to ensure that 
we only explore unique counterexample traces.

\paragraph*{\textbf{Systematic exploration of counterexamples}}

The order of exploring counterexamples is crucial to satisfy {\em monotonicity}, 
i.e., to reduce the channel capacity (a standard metric to quantify the information 
flow from sensitive input to attacker observation) of $\program$ with each round of 
patch generation. To this end, \CC employs a strategy that can be visualized 
as an exploration of the equivalence classes of observations ({\em e.g.} \#cache misses), 
i.e., we explore all counterexamples in the same equivalence class in one shot. 
In order to find another counterexample exhibiting the same 
observation as observation $o$, we modify the verification goal as follows, 
for timing and trace-based attacks, respectively ({\em cf.} \autoref{eq:xi} for $\xi$): 
%
\begin{equation}
\label{eq:refinement-same-equivalence-class}
\begin{gathered}
\Psi  \wedge \neg \left ( \left ( \sum_{i=1}^{N} miss_i \right ) \ne o \right );
\\
\Psi  \wedge \neg  
\left (
    \begin{array}[b]{c}
      \left | dom(\xi) \right | \ne \left [ \left | dom(\xi) \right | \right ]_o \\
      \vee \ \parallel_{r_i \in dom(\xi)} \xi(r_i) \ne \left [ \parallel_{r_i \in dom(\xi)} \xi(r_i) \right ]_o
    \end{array} 
\right ) 
\end{gathered}
\end{equation}
where $[X]_o$ captures the valuation of $X$ with respect to observation $o$ 
and $N$ is the total number of symbolically executed, memory-related 
instructions.
If \autoref{eq:refinement-same-equivalence-class} is unsatisfiable, then it 
captures the absence of any more counterexample with observation $o$. 
We note that $\Psi$ is automatically refined to avoid discovering duplicate 
or spurious counterexamples. 
If \autoref{eq:refinement-same-equivalence-class} is satisfiable, our 
checker provides another real counterexample with the observation 
$o$. We repeat the process until no more real counterexample with 
the observation $o$ is found, at which point 
\autoref{eq:refinement-same-equivalence-class} becomes unsatisfiable. 
%


To explore a different equivalence class of observation than that of 
observation $o$, \CC negates the verification goal. For 
$\objective_{time}$, as an example, the verification goal is changed as follows:
{
\begin{equation}
\label{eq:refinement-diff-equivalence-class}
\begin{split}
\Psi \wedge \neg \left ( \left ( \sum_{i=1}^{N} miss_i \right ) = o \right )
\end{split}
\end{equation}}
We note that \autoref{eq:refinement-diff-equivalence-class} is satisfiable 
if and only if there exists an execution trace with observation differing 
from $o$. 
%

\paragraph*{\textbf{Runtime actions to improve side-channel freedom}}
\label{sec:patching}
Our checker maintains the record of all explored observations and the symbolic 
conditions capturing the equivalence classes of respective observations. 
At each round of patch (i.e. runtime action) synthesis, we walk through this 
record and compute the necessary runtime actions for improving cache side-channel 
freedom. 

\paragraph*{\textbf{$\mathbf{\objective_{time}}$}}
Assume $\Omega = \{\langle \nu_1, o_1 \rangle, \langle \nu_2, o_2 \rangle, \ldots, \langle \nu_k, o_k \rangle\}$ 
where each $o_i$ captures a unique number of observed cache misses and $\nu_i$ 
symbolically captures all inputs that lead to observation $o_i$. Our goal 
is to manipulate executions so that they lead to the same number of cache 
misses. To this end, the patch synthesis stage determines the amount of 
cache misses that needs to be added for each element in $\Omega$. Concretely, 
the set of runtime actions generated are as follows: 
{
\begin{equation}
\label{eq:patch-time}
\begin{split}
\left \langle \nu_1, \left ( \max_{i \in [1,k]}o_i - o_1 \right ) \right \rangle, 
\ldots, 
\left \langle \nu_k,  \left ( \max_{i \in [1,k]}o_i - o_k \right ) \right \rangle
\end{split}
\end{equation}}
In practice, when a program is run with input $I$, we check whether 
$I \in \nu_x$ for some $x \in [1,k]$. Subsequently,  
$\left ( \max_{i \in [1,k]} o_i - o_x \right )$ cache misses were 
injected before the program starts executing.

\paragraph*{\textbf{$\mathbf{\objective_{trace}}$}}
During trace-based attacks, the attacker makes an observation on the sequence 
of cache hits and misses in an execution trace. 
%
Therefore, our goal is to manipulate executions in such a fashion that all 
execution traces lead to the same sequence of cache hits and misses. 
To accomplish this, each runtime action involves the injection of cache 
misses or hits before execution, after execution or at an arbitrary point 
of execution. It also involves invalidating an address in cache. 
Concretely, this is formalized as follows: 
\begin{equation}
\label{eq:patch-trace}
\left \langle \nu_i, \left \langle (c_1, a_1), (c_2, a_2), \ldots , (c_k, a_k) \right \rangle \right \rangle
\end{equation}
where $\nu_i$ captures the symbolic input condition where the runtime actions are employed. 
For any input satisfying $\nu_i$, we count the number of instructions executed. 
If the number of executed instructions reaches $c_j$, then we perform 
the action $a_j$ ({\em e.g.} injecting hits/misses or invalidating an 
address in cache), for any $j \in [1,k]$. 

As an example, consider a trace-based attack in the example of 
\autoref{fig:example}(b). Our checker will manipulate counterexample 
traces by injecting cache misses and hits as follows (injected cache 
hits and misses are highlighted in bold): 
\begin{equation*}
tr''_1 \equiv \langle \mathbf{miss}, miss, hit, hit \rangle;\ \ 
tr''_2 \equiv \langle miss, miss, hit, \mathbf{hit} \rangle
\end{equation*}
Therefore, the following actions are generated to ensure cache side-channel 
freedom against trace-based attacks: 
\begin{equation*}
\left \langle key=255, \left \langle (0, miss) \right \rangle \right \rangle,
\left \langle 0 \leq key \leq 254, \left \langle (3, hit) \right \rangle \right \rangle
\end{equation*}
We use string alignment algorithm~\cite{smw-url} to make two traces 
equivalent (via insertion of cache hits/misses or substitution of 
hits to misses). 


\paragraph*{\textbf{Practical consideration}}
In practice, the injection of a cache miss can be performed via accessing 
a fresh memory block ({\em cf.} \autoref{fig:example}(d)). 
However, unless the injection of a cache miss happens to be in the beginning 
or at the end of execution, the cache needs to be disabled before and enabled 
after such a cache miss. 
Consequently, our injection of misses does not affect 
cache states. 
In ARM-based processor, this is accomplished  
via manipulating the \texttt{C} bit of \texttt{CP15} register.
%
%
The injection of a cache hit can be performed via tracking the last accessed 
memory address and re-accessing the same address. 

To change a cache hit to a cache miss, the accessed memory address needs to 
be invalidated in the cache. \CC symbolically tracks the 
memory address accessed at each memory-related instruction. 
When the program is run with input $I \in \nu_i$, we concretize all memory 
addresses with respect to $I$. Hence, while applying an action that involves 
cache invalidation, we know the exact memory address that needs to be invalidated. 
In ARM-based processor, the instruction \texttt{MCR} provides capabilities 
to invalidate an address in the cache. 

We note the preceding manipulations on an execution requires additional 
registers. We believe this is possible by using some system register or using a 
locked portion in the cache.

\paragraph*{\textbf{Properties guaranteed by \CC}}

\CC satisfies the following crucial properties (proofs are included 
in the appendix) on channel capacity, shannon 
entropy and min entropy; which are standard metrics to quantify the 
information flow from sensitive inputs to the attacker observation. 

\begin{app}
\label{prop:monotone}
{\textbf{(Monotonicity)}} 
Consider a victim program $\program$ with sensitive input $\mathcal{K}$. 
Given attack models $\objective_{time}$ or $\objective_{trace}$, assume 
that the channel capacity to quantify the uncertainty of guessing $\mathcal{K}$ 
is $\mathcal{G}_{cap}^{\program}$. \CC guarantees that $\mathcal{G}_{cap}^{\program}$ 
monotonically decreases with each synthesized patch 
(cf. \autoref{eq:patch-time}-\ref{eq:patch-trace}) employed at runtime. 
\end{app}

\begin{app}
\label{prop:converge}
{\textbf{(Convergence)} 
Consider a victim program $\program$ with sensitive input $\mathcal{K}$. 
In the absence of any attacker, assume that the uncertainty to guess 
$\mathcal{K}$ is $\mathcal{G}_{cap}^{init}$, $\mathcal{G}_{shn}^{init}$ and 
$\mathcal{G}_{min}^{init}$, via channel capacity, Shannon entropy and Min 
entropy, respectively. If \CC terminates and all synthesized patches 
are applied at runtime, 
then the channel capacity 
(respectively, Shannon entropy and Min entropy) 
will remain $\mathcal{G}_{cap}^{init}$ 
(respectively, $\mathcal{G}_{shn}^{init}$ and $\mathcal{G}_{min}^{init}$) 
even in the presence of attacks captured via $\objective_{time}$
and $\objective_{trace}$.
}
\end{app}

\section{Implementation and Evaluation}
\label{sec:evaluation}


\paragraph*{\textbf{Implementation setup}}
The input to \CC is the target program and a cache configuration. 
We have implemented \CC on top of \texttt{CBMC} bounded model checker~\cite{cbmc-url}. 
It first builds a formula representation 
of the input program via symbolic execution. Then, it checks the 
(un)satisfiability of this formula against a specification property.
%
Despite being a bounded model checker, CBMC is used as a classic 
verification tool in our experiments. In 
particular for program loops, CBMC first attempts to derive loop bounds 
automatically. If CBMC fails to derive certain loop bounds, then the respective 
loop bounds need to be provided manually. Nevertheless, during the verification 
process, CBMC checks all manually provided loop bounds and the verification 
fails if any such bound was erroneous. In our experiments, all loop bounds 
were automatically derived by CBMC. In short, if \CC successfully verifies 
a program, then the respective program exhibits cache side-channel freedom 
for the given cache configuration and targeted attack models.

The implementation of our checker impacts the entire workflow of \texttt{CBMC}. 
We first modify the symbolic execution engine of \texttt{CBMC} to insert the 
predicates related to cache semantics. As a result, upon the termination of 
symbolic execution, the formula representation of the program encodes both the 
cache semantics and the program semantics. Secondly, we systematically rewrite 
this formula based on our abstraction refinement, with the aim of verifying 
cache side-channel freedom. Finally, we modify the verification engine of 
\texttt{CBMC} to systematically explore different counterexamples, instrument 
patches and refining the side-channel freedom properties {\em on-the-fly}. 
To manipulate and solve symbolic formulas, we leverage \texttt{Z3} theorem 
prover. 
All reported experiments were performed on an Intel I7 machine, 
having 8GB RAM and running OSX. 

\begin{table*}
 \begin{center}
  \caption{Summary of \CC Evaluation. Timeout is set to ten minutes. $\mathit{Time}$ captures the 
  time taken by \CC (i.e. \# predicates $\left |  \mathit{Pred}_{cur} \right |$), whereas 
  $\mathit{Time}_{all}$ captures the time taken when all predicates 
  (i.e. \# predicates $\left |  \mathit{Pred}_{set} \cup \mathit{Pred}_{tag} \right |$) are 
  considered.}
  {\scriptsize 
  \begin{tabular}{|c||c|c|c|c|c|c|c|c|c|c|c|}
  \hline
  Library & Routine & Size &  & 
  \multicolumn{4}{|c|}{$\mathbf{\objective_{time}}$} & 
  \multicolumn{4}{|c|}{$\mathbf{\objective_{trace}}$}\\
  \cline{5-6} \cline{7-12}
          &   & (LOC) &  $\left | \mathit{Pred}_{set} \cup \mathit{Pred}_{tag} \right |$ 
          & $\left |  \mathit{Pred}_{cur} \right |$ & Result & $\mathit{Time}$ & $\mathit{Time}_{all}$  
          & $\left |  \mathit{Pred}_{cur} \right |$ & Result & $\mathit{Time}$ & $\mathit{Time}_{all}$\\
          &   &   &  &  &  & (secs) & (secs) &  &  & (secs) & (secs)\\
  \hline	 
   & \texttt{AES128} & 740 & 231959  & 115537 & \xmark  & 148.73 & {\bf timeout} & 115545 & \xmark & 175.34 & {\bf timeout}\\
  \cline{2-12}
  \texttt{OpenSSL} & \texttt{DES} & 2124 &  205567 & 95529 & \xmark & 105.87 & {\bf timeout} & 95639 & \xmark & 110.32 & 
  {\bf timeout}\\
  \cline{2-12}
  \cite{openssl-url} & \texttt{RC5} & 1613 & 50836 & 25277 & \cmark & 26.88 &  541.35 &  26277 & \cmark & 34.84 & 585.32\\
  \hline
  \hline
  \texttt{GDK} & \texttt{keyname} & 712 & 20827 & 18178  & \xmark & 21.75 &  120.21 & 18178 & \xmark & 24.32  & 125.11\\
  \cline{2-12}
  \cite{gdk-url} & \texttt{unicode} & 862 & 21917 & 19178  & \xmark & 27.49 & 117.78 & 19178 & \xmark & 32.09 & 119.56\\
  \hline	
  \hline
    & \texttt{fix\_eq} & 334 &  71 & 32 & \cmark & 1.70 & 5.70 & 32 & \cmark & 4.31 & 6.08\\
	\cline{2-12}
	& \texttt{fix\_cmp} & 400 &  957	& 474 & \cmark & 2.09 & 6.12 & 476 & \cmark & 2.08 & 6.15\\
	\cline{2-12}
	& \texttt{fix\_mul} & 930 & 33330 & 16651  & \cmark & 22.44 & 100.32 & 17016 & \cmark & 20.19 & 121.37\\
	\cline{2-12}
	& \texttt{fix\_conv\_64} & 350 &  211 & 102 & \cmark & 1.69 & 1.81 & 102 & \cmark & 1.78  & 1.98\\
	\cline{2-12}
	& \texttt{fix\_sqrt} & 2480 & 150127	& 74961  & \cmark & 60.54 & 359.94 & 87834 & \cmark & 67.90 & 581.45\\
	\cline{2-12}
\texttt{fixedt} & \texttt{fix\_exp} & 1128 & 101418 & 50655 & \cmark & 53.47 &  400.11 & 56194 & \cmark & 55.12 & 416.21\\
	\cline{2-12}
\cite{arithmetic-url} & \texttt{fix\_ln} & 1140 & 92113 & 46025  & \cmark & 58.89 & 114.89 & 47065 & \cmark & 61.11 & 127.21\\
	\cline{2-12}
	& \texttt{fix\_pow} & 2890 & 643961 & 321885  & \cmark & 389.97 & {\bf timeout} & 323787 & \cmark & 377.55 & {\bf timeout} \\
	\cline{2-12}
	& \texttt{fix\_ceil} & 390 &  266 & 128 & \cmark & 1.70 & 1.70 &  128 & \cmark & 4.65 & 4.70\\
	\cline{2-12}
	& \texttt{fix\_conv\_double} & 650 & 1921 & 953  & \cmark & 2.70 & 2.75 & 945 & \cmark & 4.16 & 4.20\\
	\cline{2-12}
	& \texttt{fix\_frac} & 370 & 60172 & 53638  & \xmark & 22.14 & 23.18 & 51432 & \xmark & 22.15 & 24.58\\
    \hline
    \hline
    & \texttt{eccmadd} & 1550 &  324661 & 162058 & \cmark & 220.59 & 437.77 & 165453 & \cmark & 214.16 & 502.94\\
	\cline{2-12}
	& \texttt{eccnorm} & 1303 & 165291 & 82464  & \cmark & 105.97 & 198.90 & 83100 & \cmark & 114.93 & 219.09\\
	\cline{2-12}
	& \texttt{pt\_setup} & 1345 &  3850 & 1866 & \cmark & 10.45 & 11.66 & 1901 & \cmark & 14.90 & 14.95\\
	\cline{2-12}
	& \texttt{eccdouble} & 1364 & 531085	& 265219  & \cmark & 285.70 & 558.78  & 267889 & \cmark & 312.31 & 597.18\\
	\cline{2-12}
	& \texttt{R1\_to\_R2} & 1352 &  85750 & 42731 & \cmark & 73.07 &  249.12 & 41014 & \cmark & 84.11 & 398.77\\
	\cline{2-12}
\texttt{FourQ} & \texttt{R1\_to\_R3} & 1328 &  25246	& 12538 & \cmark & 26.95 &  53.21 & 12555 & \cmark & 24.89 & 54.45\\
	\cline{2-12}
\cite{fourqlib-url} & \texttt{R2\_to\_R4} & 1322 & 28415	& 14105 & \cmark & 32.29 & 55.11 & 17001 & \cmark & 31.12 & 61.88\\
	\cline{2-12}
	& \texttt{R5\_to\_R1} & 1387 & 12531	 & 5255  & \cmark & 22.07 & 34.05 & 5278 & \cmark & 24.32 & 52.88 \\
	\cline{2-12}
	& \texttt{eccpt\_validate} & 1406 &  57129 & 38012 & \xmark & 34.48 & 61.91  & 37948 & \xmark & 34.31  & 71.54 \\
    \hline
  \end{tabular}}
\label{tab:efficiency}   
\end{center}
\end{table*}

\begin{table*}
 \begin{center}
  \caption{Overhead in generating monitors and due to the applied runtime actions. $\mathit{Time}$ captures 
  the time to generate all patches. The overhead (maximum and average) captures the number of extra cache misses, 
  hits and invalidations introduced by \CC in absolute term (i.e. \# actions) and with respect to the total number 
  of instructions (i.e. \% actions). }
  {\scriptsize 
  \vspace*{-0.1in}
  \begin{tabular}{|c||c|c|c|c|c|c|c|c|c|c|c|c|}
  \hline
  Routine & \multicolumn{6}{|c|}{$\mathbf{\objective_{time}}$} & 
  \multicolumn{6}{|c|}{$\mathbf{\objective_{trace}}$}\\
  \cline{2-3} \cline{4-13}
          & \#Equivalence  & $\mathit{Time}$ & \multicolumn{2}{|c|}{Max. overhead} 
          & \multicolumn{2}{|c|}{Avg. overhead}
          & \#Equivalence & $\mathit{Time}$ & \multicolumn{2}{|c|}{Max. overhead} 
          & \multicolumn{2}{|c|}{Avg. overhead}\\
          \cline{4-7}\cline{10-13}
          & class & (secs) & (\# actions) & (\% actions) & (\# actions) & (\% actions) 
          & class & (secs) & (\# actions) & (\% actions) & (\# actions) & (\% actions) \\
  \hline	 
  \texttt{AES128} & 3 & 5 & 117 & 0.1\% & 60 & 0.05\% & 38 & 1271 & 120 & 0.1\% & 90 & 0.07\% \\
  \cline{2-13}
  \texttt{DES} & 333 & 444 & 300 & 0.3\% & 150 & 0.15\% & 982 & 12601 & 371 & 0.6\% & 231 & 0.4\% \\
  \cline{2-13}
  \texttt{fix\_frac} & 82 & 521 & 80 & 1.2\% & 40 & 0.6\% & 82 & 956 & 119 & 1.5\% & 62 & 1.1\% \\
  \cline{2-13}
  \texttt{eccpt\_validate} & 20 & 22 & 18 & 0.09\% & 9 & 0.05\% & 20 & 234 & 21 & 1.1\% & 11 & 0.04\% \\
  \cline{2-13}
  \texttt{keyname} & 41 & 1119  & 39 & 2.6\% & 19 & 1.2\% & 41 & 1324 & 45 & 3.2\% & 28 & 2.1\% \\
  \cline{2-13}
  \texttt{unicode} & 43 & 197 & 41 & 2.4\% & 20 & 1.2\% & 43 & 229 & 49 & 2.4\% & 28 & 1.8\% \\
  \hline
  \end{tabular}}
\label{tab:monitor}   
\end{center}
\end{table*}

\paragraph*{\textbf{Subject programs and cache}}
We have chosen security-critical subjects from \texttt{OpenSSL}~\cite{openssl-url}, 
\texttt{GDK}~\cite{gdk-url}, arithmetic routines 
from \texttt{libfixedtimefixedpoint}~\cite{arithmetic-url} 
and elliptic curve routines from \texttt{FourQlib}~\cite{fourqlib-url} 
to evaluate \CC (cf. \autoref{tab:efficiency}). 
We include representative routines exhibiting constant 
cache-timing, as well as routines exhibiting variable cache timing. 
We set the default cache to be 1KB direct-mapped, with a line size of 32 bytes.

\paragraph*{\textbf{Efficiency of checking}}
\autoref{tab:efficiency} captures a summary of our evaluation for \CC. 
The outcome of this evaluation is either a successful verification (\cmark) 
or a non-spurious counterexample (\xmark). 
\CC accomplished the verification tasks for all subjects only 
within a few minutes.
The maximum time taken by our checker was $390$ seconds for the routine 
\texttt{fix\_pow} -- a constant time implementation of powers ($x^y$). 
\texttt{fix\_pow} has complex memory access patterns, however, its 
flat structure ensures cache side-channel freedom. 

To check the effectiveness of our abstraction refinement process, 
we compare \CC with a variant of our checker where all predicates in 
$\mathit{Pred}_{set} \cup \mathit{Pred}_{tag}$ are considered. Therefore, 
such a variant does not employ any abstraction refinement, as the set of 
predicates $\mathit{Pred}_{set} \cup \mathit{Pred}_{tag}$ is sufficient 
to determine the cache behaviour of all instructions in the program. 
We compare \CC with this variant in terms of the number of predicates, 
as well as the verification time. We record the set of predicates 
$\mathit{Pred}_{cur}$ considered in \CC when it terminates with a successful 
verification (\cmark) or a real counterexample (\xmark). 
\autoref{tab:efficiency} clearly demonstrates the effectiveness of our 
abstraction refinement process. Specifically, for \texttt{AES}, \texttt{DES} 
and \texttt{fix\_pow}, the checker does not terminate in {\em ten minutes} 
when all predicates in $\mathit{Pred}_{set} \cup \mathit{Pred}_{tag}$ are 
considered during the verification process. In general, the refinement 
process reduces the number of considered predicates by a factor of $1.81x$ on 
average. This leads to a substantial improvement in verification time, 
as observed from \autoref{tab:efficiency}.

The routines chosen from \texttt{OpenSSL} library are single path programs. 
However, \texttt{AES} and \texttt{DES} exhibit input-dependent memory 
accesses, hence, violating side-channel freedom. 
The other routines violate cache side-channel freedom due to input-dependent 
loop trip counts. For example, routines chosen from the \texttt{GDK} 
library employ a binary search of the input keystroke over a large table.  
We note that both \texttt{libfixedtimefixedpoint} and \texttt{FourQlib} 
libraries include comments involving the security risks 
in \texttt{fix\_frac} and \texttt{ecc\_point\_validate} 
({\em cf.} \texttt{validate} in \autoref{tab:efficiency}).
%
\begin{note}
{\em \CC verifies or generates real counterexamples at an average 
within 70.38 secs w.r.t. attack model $\objective_{time}$ and at an average 
within 74.12 secs w.r.t. $\objective_{trace}$. Moreover, the abstraction 
refinement process embodied within \CC substantially reduces the verification 
time as opposed to when no refinement process was employed.
}
\end{note}

\paragraph*{\textbf{Overhead from monitors}}
We evaluated the time taken by \CC for counterexample exploration 
and patch generation ({\em cf.} \autoref{sec:monitor}). 
For each generated patch, we have also evaluated the 
overhead induced by the same at runtime (cf. \autoref{tab:monitor}).
%
%

\autoref{tab:monitor} captures the maximum and average overhead 
induced by \CC at runtime. We compute the overhead via the number of 
additional runtime actions (i.e. cache misses, hits or invalidations) 
introduced solely via \CC. In absolute terms, the maximum (average) 
overhead captures the maximum (average) number of runtime actions induced 
over all equivalence classes. 
The maximum overhead was introduced in case of \texttt{DES} -- 300 actions 
for $\objective_{time}$ and 371 actions for $\objective_{trace}$. This is 
primarily due to the difficulty in making a large number of traces 
equivalent in terms of the number of cache misses and the sequence of 
hit/miss, respectively. Although the number of actions introduced by 
\CC is non-negligible, we note that their effect is minimal on the 
overall execution. To this end, we execute the program for 100 different 
inputs in each explored equivalence class and measure the overhead 
introduced by \CC (cf. ``\% actions" in \autoref{tab:monitor}) with 
respect to the total number of instructions executed. We observe that 
the maximum overhead reaches up to 3.2\% and the average overhead is 
up to 2.1\%. We believe this overhead is acceptable in the light of 
cache side-channel freedom guarantees provided by \CC.

Except \texttt{AES} and \texttt{DES}, 
the cache behaviour of a single program path is independent of 
program inputs. 
For the respective subjects, exactly the same number of equivalence classes 
were explored for both attack models (cf. \autoref{tab:monitor}). Each 
explored equivalence class was primarily attributed to a unique program path. 
Nevertheless, due to more involved computations ({\em cf.} \autoref{sec:monitor}), 
the overhead of \CC in attack model $\objective_{trace}$ is higher than the 
overhead in attack model $\objective_{time}$. 
%
%
%
As observed from \autoref{tab:monitor}, \CC discovers significantly more 
equivalence classes w.r.t. attack model 
$\objective_{trace}$ as compared to the number of equivalence classes 
w.r.t. attack model $\objective_{time}$. This implies \texttt{AES} is more 
vulnerable to $\objective_{trace}$ as compared to $\objective_{time}$. 
Excluding \texttt{DES} subject to $\objective_{trace}$, our exploration 
terminates in all scenarios within $20$ mins. 


%

\begin{note}
{\em To explore counterexample and generate patches, \CC takes  
$2.27$x more time with $\objective_{trace}$, as 
compared to $\objective_{time}$. Moreover, excluding 
\texttt{DES}, \CC explores all equivalence classes of observations 
within 20 minutes in all scenarios. Finally, the runtime overhead 
induced by \CC is only up to 3.2\% with respect to the number of 
executed instructions.
}
\vspace{-0.2cm}
\end{note}

\paragraph*{\textbf{Sensitivity w.r.t. cache configuration}}
We evaluated \CC for a variety of cache associativity 
(1-way, 2-way and 4-way), cache size (from 1KB to 8KB) and with 
LRU as well as FIFO replacement policies (detailed experiments are 
included in the appendix). 
%
We observed that the verification time increases marginally (about 7\%) 
when set-associative caches were used instead of direct-mapped caches 
%
and does not vary significantly with respect to replacement policy. Finally, we observed 
changes in the number of equivalence classes of observations for both 
\texttt{AES} and \texttt{DES} while running these subjects with different 
replacement policies. However, neither \texttt{AES} nor \texttt{DES} 
satisfied cache side-channel freedom for any of the cache size and 
replacement policies tested in our evaluation. 
The relatively low verification time results from the fact that the 
total number of predicates (i.e. $\mathit{Pred}_{set} \cup \mathit{Pred}_{tag}$) 
is independent of cache size and replacement policy. Nevertheless, 
the symbolic encoding for set-associative caches is more involved 
than direct-mapped caches. This results in an average increase to 
the number of predicates considered for verification 
(i.e. $\mathit{Pred}_{cur}$) by a factor of $1.5x$. 
However, such an increased number of predicates does not translate to 
significant verification timing for set-associative caches. 


\section{Review of Prior Works}
\label{sec:related-work}
Earlier works on cache analysis are based on abstract interpretation~\cite{ferdinand-rts-paper} 
and its combination with model checking~\cite{aimc-paper}, 
to estimate the worst-case execution time (WCET) of a program. 
In contrast to these approaches, \CC automatically builds and refine the abstraction 
of cache semantics for verifying side-channel freedom. 
%
Cache attacks are one of the most critical side-channel 
attacks~\cite{side-channel-survey-paper,timing-attack-paper,trace-attack-paper,cachegames,cache_template_paper,cachebleed-attack,flush-and-reload,gernot-cache-attack-paper,cache-storage-paper}. 
In contrast to the literature on side-channel attacks, we 
do not engineer new cache attacks in this paper. Based on a 
configurable attack model, \CC verifies and reinstates the cache side-channel 
freedom of arbitrary programs.

%
Orthogonal to approaches proposing countermeasures~\cite{rubylee:rpcache,ndss15-paper-diversity}, 
the fixes generated by \CC is guided by program verification output. Thus, 
\CC can provide cache side-channel freedom guarantees about the fixed program.
%
%
Moreover, \CC can be leveraged to formally 
verify whether existing countermeasures are capable to ensure 
side-channel freedom.

In contrast to recent approaches on statically analyzing cache side 
channels~\cite{cav12-paper,cacheaudit,post17-paper,adaptive-side-channel},  
our \CC approach automatically constructs and refines the 
abstractions for verifying cache side-channel freedom. Moreover, contrary 
to \CC, approaches 
based on static analysis are not directly applicable when the underlying 
program does not satisfy cache side-channel freedom. 
\CC targets verification of arbitrary software programs, over 
and above constant-time implementations~\cite{ccs14-paper,usenix16-paper}.
Existing works based on symbolic execution~\cite{max_smt_side_channel_paper,qip-paper}, 
taint analysis~\cite{taint-tracking-paper,taint-analysis} and verifying timing-channel 
freedom~\cite{pldi17-decomposition} ignore cache attacks. Moreover, these works do not 
provide capabilities for automatic abstraction refinement and patch synthesis for ensuring 
side-channel freedom. 
Finally, in contrast to these works, we show that our \CC approach scales 
with routines from real cryptographic libraries.

Finally, recent approaches on testing and quantifying cache side-channel 
leakage~\cite{tacas17-paper,icstw-paper,memocode17-paper} 
are complementary to \CC. These works have the flavour of testing 
and 
and they do not provide capabilities to ensure cache side-channel freedom. 





\section{Discussion}
\label{sec:discussion}

In this paper, we propose \CC, a novel approach to automatically 
verify and restore cache side-channel freedom of arbitrary programs. 
The key novelty in our approach is two fold. Firstly, our \CC 
approach automatically builds and refines abstraction of cache 
semantics. 
Although targeted to verify cache side-channel freedom, 
we believe \CC is applicable to verify other cache timing properties, 
such as WCET.
Secondly, the core symbolic engine of \CC systematically combines its reasoning 
power with runtime monitoring to ensure cache side-channel freedom 
during program execution. 
Our evaluation reveals promising results, for 25 routines from 
several cryptographic libraries, 
\CC (dis)proves cache side-channel freedom within an average 75 
seconds. 
Moreover, in most scenarios, \CC 
generated patches within 20 minutes to ensure cache side-channel 
freedom during program execution. 
Despite this result, we believe 
that \CC is only an initial step for the automated verification of 
cache side-channel freedom. In particular, we 
do not account cache attacks that are more powerful than timing or 
trace-based attacks. Besides, we do not implement the synthesized 
patches in a commodity embedded system to check their performance 
impact. We hope that the community will take this effort forward 
and push the adoption of formal tools for the evaluation of cache 
side-channel. For reproducibility and research, our tool and all 
experimental data are publicly available: 
\textbf{(blinded)}

{
\bibliographystyle{ACM-Reference-Format}
\bibliography{cache_checker}}


\section*{Appendix}
The appendix includes additional cache models (e.g. LRU and FIFO) incorporated 
within \CC, the theoretical guarantees and additional experimental results. 

\subsection*{Theoretical Guarantees}
In this section, we include the detailed proof of the properties satisfied by \CC.

\begin{app}
\label{prop:monotone-appendix}
{\textbf{(Monotonicity)}} 
Consider a victim program $\program$ with sensitive input $\mathcal{K}$. 
Given attack models $\objective_{time}$ or $\objective_{trace}$, assume 
that the channel capacity to quantify the uncertainty of guessing $\mathcal{K}$ 
is $\mathcal{G}_{cap}^{\program}$. \CC guarantees that $\mathcal{G}_{cap}^{\program}$ 
monotonically decreases with each synthesized patch 
(cf. \autoref{eq:patch-time}-\ref{eq:patch-trace}) employed at runtime. 
\end{app}

\begin{proof}
%
Consider the generic attack model $\objective: \{h,m\}^{*} \rightarrow \mathbb{X}$ 
that maps each trace to an element in the countable set $\mathbb{X}$. 
For a victim program $\program$, assume $\mathit{TR} \subseteq \{h,m\}^{*}$ is 
the set of all execution traces. After one round of patch synthesis, assume 
$\mathit{TR}' \subseteq \{h,m\}^{*}$ is the set of all execution traces in 
$\program$ when the synthesized patches are applied at runtime. By construction,  
each round of patch synthesis merges two equivalence classes of observations 
({\em cf.} Algorithm~\ref{alg:monitor}), hence, making them indistinguishable 
by the attacker $\objective$. As a result, the following relationship holds:
{
\begin{equation}
\label{eq:trace-neutralize}
\left | \objective (\mathit{TR}) \right | = \left | \objective (\mathit{TR}') \right | + 1
\end{equation} }
Channel capacity $\mathcal{G}_{cap}^{\program}$  equals to 
$\log|\objective (\mathit{TR})|$ for the original 
program $\program$, but it reduces to $\log|\objective \left ( \mathit{TR}' \right )|$ when 
the synthesized patches are applied. We conclude the proof as the same argument holds for 
any round of patch synthesis. 
\end{proof}

\begin{app}
\label{prop:converge-appendix}
{\textbf{(Convergence)} 
Let us assume a victim program $\program$ with sensitive input $\mathcal{K}$. 
In the absence of any attacker, assume that the uncertainty to guess 
$\mathcal{K}$ is $\mathcal{G}_{cap}^{init}$, $\mathcal{G}_{shn}^{init}$ and 
$\mathcal{G}_{min}^{init}$, via channel capacity, Shannon entropy and Min 
entropy, respectively. If our checker terminates and all synthesized patches 
are applied at runtime, then our framework guarantees that the channel capacity 
(respectively, Shannon entropy and Min entropy) 
will remain $\mathcal{G}_{cap}^{init}$ 
(respectively, $\mathcal{G}_{shn}^{init}$ and $\mathcal{G}_{min}^{init}$) 
even in the presence of attacks captured via $\objective_{time}$ and 
$\objective_{trace}$.}
\end{app}

\begin{proof}
Consider the generic attack model $\objective: \{h,m\}^{*} \rightarrow \mathbb{X}$, 
mapping each execution trace to an element in the countable set $\mathbb{X}$. 
We assume a victim program $\program$ that exhibits a set of execution traces 
$\mathit{TR} \subseteq \{h,m\}^{*}$. 
From \autoref{eq:trace-neutralize}, we know that $|\objective (\mathit{TR})|$ 
decreases with each round of patch synthesis. 
Given that our checker terminates, we obtain the program $\program$, together with 
a set of synthesized patches that are applied when $\program$ executes. 
Assume $\mathit{TR}^{f} \in \{h,m\}^{*}$ is the set of execution traces obtained 
from $\program$ when all patches are systematically applied. 
Clearly, $\left | \objective \left ( \mathit{TR}^{f} \right ) \right | = 1$. 

The channel capacity of $\program$, upon the termination of our checker, is 
$\log \left (\left | \objective \left ( \mathit{TR}^{f} \right ) \right | \right )$ = 
$\log 1$ = $0$. This concludes that the channel capacity does not change even 
in the presence of attacks $\objective_{time}$ and $\objective_{trace}$.

For a given distribution $\lambda$ of sensitive input $\mathcal{K}$, Shannon 
entropy $\mathcal{G}_{shn}^{init}$ is computed as follows: 
\begin{equation}
\mathcal{G}_{shn}^{init} (\lambda) = - \sum_{\mathcal{K} \in \mathbb{K}} \lambda(\mathcal{K}) \log_{2} \lambda(\mathcal{K})
\end{equation}
where $\mathbb{K}$ captures the domain of sensitive input $\mathcal{K}$. For 
a given equivalence class of observation $o \in \objective \left ( \mathit{TR}^{f} \right )$, 
the remaining uncertainty is computed as follows: 
\begin{equation}
\label{eq:shannon-remaining-one-appendix}
\mathcal{G}_{shn}^{final} (\lambda_o) = - \sum_{\mathcal{K} \in \mathbb{K}} \lambda_o(\mathcal{K}) \log_{2} \lambda_o(\mathcal{K})
\end{equation}
$\lambda_o(\mathcal{K})$ captures the probability that the sensitive input is $\mathcal{K}$, 
given the observation $o$ is made by the attacker. Finally, to evaluate the 
remaining uncertainty of the patched program version, $\mathcal{G}_{shn}^{final} (\lambda_o)$ 
is averaged over all equivalence class of observations as follows:  
\begin{equation}
\label{eq:shannon-remaining-average-appendix}
\mathcal{G}_{shn}^{final} \left ( \lambda_{\objective \left ( \mathit{TR}^{f} \right )} \right ) 
= \sum_{o \in \objective \left ( \mathit{TR}^{f} \right )} pr(o)\ \mathcal{G}_{shn}^{final} (\lambda_o)
\end{equation}
where $pr(o)$ captures the probability of the observation $o \in \objective \left ( \mathit{TR}^{f} \right )$. 
However, we have $\left | \objective \left ( \mathit{TR}^{f} \right ) \right | = 1$. Hence, for any 
$o \in \objective \left (\mathit{TR}^{f} \right )$, we get $pr(o)=1$ and $\lambda_o(\mathcal{K})=\lambda(\mathcal{K})$. 
Plugging these observations into \autoref{eq:shannon-remaining-average-appendix} and \autoref{eq:shannon-remaining-one-appendix}, 
we get the following: 
\begin{equation}
\label{eq:proof-shannon-appendix}
\boxed{
\begin{split}
\mathcal{G}_{shn}^{final} \left ( \lambda_{\objective \left ( \mathit{TR}^{f} \right ) } \right ) 
&= \sum_{o \in \objective \left ( \mathit{TR}^{f} \right )} \mathcal{G}_{shn}^{final} (\lambda_o) 
\\
&= - \sum_{\mathcal{K} \in \mathbb{K}} \lambda(\mathcal{K}) \log_{2} \lambda(\mathcal{K}) 
\\
&= \mathcal{G}_{shn}^{init} (\lambda)
\end{split}}
\end{equation}

Finally, for a given distribution $\lambda$ of sensitive input $\mathcal{K}$, 
the min entropy $\mathcal{G}_{min}^{init}$ is computed as follows:
\begin{equation}
\label{eq:min-entropy-appendix}
\mathcal{G}_{min}^{init} (\lambda) = - \log_{2}\ \max_{\mathcal{K} \in \mathbb{K}}\ \lambda(\mathcal{K}) 
\end{equation}
Therefore, min entropy captures the best strategy of an attacker, that is, to 
choose the most probable secret. 

Similar to Shannon entropy, for a given equivalence class of observation 
$o \in \objective \left ( \mathit{TR}^{f} \right )$, the remaining 
uncertainty is computed as follows: 
\begin{equation}
\label{eq:min-entropy-observe-one-appendix}
\mathcal{G}_{min}^{init} (\lambda_o) = - \log_{2}\ \max_{\mathcal{K} \in \mathbb{K}}\ \lambda_o(\mathcal{K}) 
\end{equation}
$\lambda_o(\mathcal{K})$ captures the probability that the sensitive input is $\mathcal{K}$, 
given the observation $o$ is made by the attacker.

Finally, we obtain the min entropy of the patched program version via the following 
relation: 
{
\begin{equation}
\label{eq:minentropy-remaining-average-appendix}
\mathcal{G}_{min}^{final} \left ( \lambda_{\objective \left ( \mathit{TR}^{f} \right ) } \right ) = 
- \log_{2} \sum_{o \in \objective \left ( \mathit{TR}^{f} \right ) } 
pr(o)\  \max_{\mathcal{K} \in \mathbb{K}}\ \lambda_o(\mathcal{K}) 
\end{equation}}

Since $pr(o)=1$ and $\lambda_o(\mathcal{K})=\lambda(\mathcal{K})$ for any 
$o \in \objective \left ( \mathit{TR}^{f} \right )$, we get the following from 
\autoref{eq:minentropy-remaining-average-appendix} and \autoref{eq:min-entropy-observe-one-appendix}: 

\begin{equation}
\label{eq:proof-min-entropy-appendix}
\boxed{
\begin{split}
\mathcal{G}_{min}^{final}  \left ( \lambda_{\objective \left ( \mathit{TR}^{f} \right ) } \right ) 
&= - \log_{2} \sum_{o \in \objective \left ( \mathit{TR}^{f} \right )}\ \max_{\mathcal{K} \in \mathbb{K}} 
\lambda_o(\mathcal{K})
\\
&= - \log_{2}\ \max_{\mathcal{K} \in \mathbb{K}}\ \lambda(\mathcal{K}) 
\\
&= \mathcal{G}_{min}^{init} (\lambda)
\end{split}}
\end{equation}
\autoref{eq:proof-shannon-appendix} and \autoref{eq:proof-min-entropy-appendix} conclude this proof. 

\end{proof}

\subsection*{Modeling LRU and FIFO cache semantics}

To formulate the conditions for conflict misses in set-associative caches, 
it is necessary to understand the notion of cache conflict. We use the following 
definition of cache conflict to formulate $\Gamma(r_i)$:

\begin{app1}
\textbf{(Cache conflict)} {\em $r_j$ generates a cache conflict to $r_i$ if and 
only if $1 \leq j < i$, $\sigma(r_j) \ne \sigma(r_i)$ and the execution of $r_j$ 
can change the relative position of $\sigma(r_i)$ within the $set(r_i)$-state 
immediately before instruction $r_i$. }
\end{app1}
Recall that $\sigma(r_i)$ captures the memory block accessed at $r_i$ and 
$set(r_i)$ captures the cache set accessed by $r_i$. The state of a cache 
set is an ordered $\mathcal{A}$-tuple -- capturing the relative positions of 
all memory blocks within the respective cache set. For instance, 
$\langle m_1, m_2 \rangle$ captures the state of a two-associative cache 
set. The rightmost memory block ({\em i.e.} $m_2$) captures the first 
memory block to be evicted from the cache set if a block $m \notin \{m_1,m_2\}$ 
is accessed and mapped to the same cache set.

\subsubsection*{\textbf{Challenges with LRU policy}}
To illustrate the unique challenges related to set-associative caches, let us 
consider the following sequence of memory accesses in a two-way associative 
cache and with LRU replacement policy: 
$(r_1:m_1) \rightarrow (r_2:m_2) \rightarrow (r_3:m_2) \rightarrow (r_4:m_1)$. 
We assume both $m_1$ and $m_2$ are mapped to the same cache set. If the cache 
is empty before $r_1$, $r_4$ will still incur a {\em cache hit}. 
This is because, $r_4$ suffers cache conflict only once, from the memory block 
$m_2$. To incorporate the aforementioned phenomenon into our cache semantics, 
we only count cache conflicts from the closest access to a given memory block. 
Therefore, in our example, we count cache conflicts to $r_4$ from $r_3$ and 
discard the cache conflict from $r_2$. 
Formally, we introduce the following additional condition for instruction $r_j$ 
to inflict a cache conflict to instruction $r_i$.

$\mathbf{\phi_{ji}^{eqv,lru}}:$ No instruction between $r_j$ and $r_i$ accesses 
the same memory block as $r_j$. This is to ensure that $r_j$ is the closest to 
$r_i$ in terms of accessing the memory block $\sigma(r_j)$. 
We capture $\phi_{ji}^{eqv,lru}$ formally as follows: 
\begin{equation}
\label{eq:set-lru-eqv}
\phi_{ji}^{eqv,lru} \equiv \bigwedge_{j < k < i} \left ( \rho_{jk}^{tag} \vee \neg \rho_{jk}^{set} \vee \neg guard_k \right )
\end{equation} 
Hence, $r_j$ inflicts a unique cache conflict to $r_i$ only if 
$\phi_{ji}^{eqv,lru}$, $\phi_{ji}^{cnf,lru} \equiv \phi_{ji}^{cnf,dir}$, 
$\phi_{ji}^{rel,lru} \equiv \phi_{ji}^{rel,dir}$ are all satisfiable. 



\subsubsection*{\textbf{Challenges with FIFO policy}}
Unlike LRU replacement policy, the cache state does not change for a cache hit in FIFO 
replacement policy. For example, consider the following 
sequence of memory accesses in a two-way associative FIFO cache: 
$(r_1:m_1) \rightarrow (r_2:m_2) \rightarrow (r_3:m_1) \rightarrow (r_4:m_1)$. Let us 
assume $m_1$, $m_2$ map to the same cache set and the cache is empty before $r_1$. In 
this example, $r_2$ generates a cache conflict to $r_4$ even though $m_1$ is accessed 
between $r_2$ and $r_4$. This is because $r_3$ is a cache hit and it does not change 
cache states.  

In general, to formulate $\Gamma(r_i)$, we need to know whether any instruction $r_j$, 
prior to $r_i$, was a cache miss. This, in turn, is captured via $\Gamma(r_j)$. 
Concretely, $r_j$ generates a unique cache conflict to $r_i$ if all the following 
conditions are satisfied.

$\mathbf{\phi_{ji}^{cnf,fifo}}:$ If $r_j$ accesses the same cache set as $r_i$, but 
accesses a different cache-tag as compared to $r_i$ and $r_j$ suffers a cache miss. 
This is formalized as follows: 
\begin{equation}
\label{eq:fifo-dif}
\phi_{ji}^{cnf,fifo} \equiv \rho_{ji}^{tag} \wedge \rho_{ji}^{set} \wedge \Gamma(r_j)
\end{equation}

$\mathbf{\phi_{ji}^{rel,fifo}}:$ No cache miss between $r_j$ and $r_i$ access the 
same memory block as $r_i$. $\phi_{ji}^{rel,fifo}$ ensures that the relative position 
of the memory block $\sigma(r_i)$ within $set(r_i)$ was not reset between 
$r_j$ and $r_i$. $\phi_{ji}^{rel,fifo}$ is formalized as follows: 
\begin{equation}
\label{eq:set-fifo-rel}
\phi_{ji}^{rel,fifo} \equiv \bigwedge_{j < k < i} \left ( \rho_{ki}^{tag} \vee \neg \rho_{ki}^{set} \vee \neg guard_k 
\vee \neg \Gamma(r_k) \right )
\end{equation} 

$\mathbf{\phi_{ji}^{eqv,fifo}}:$ No cache miss between $r_j$ and $r_i$ access the same memory 
block as $r_j$. We note that $\phi_{ji}^{eqv,fifo}$ ensures that $r_j$ is the closest cache 
miss to $r_i$ accessing the memory block $\sigma(r_j)$. This, in turn, ensures that 
we count the cache conflict from memory block $\sigma(r_j)$ to instruction $r_i$ 
{\em only once}. We formulate $\phi_{ji}^{eqv,fifo}$ as follows: 
\begin{equation}
\label{eq:set-fifo-eqv}
\phi_{ji}^{eqv,fifo} \equiv \bigwedge_{j < k < i} \left ( \rho_{jk}^{tag} \vee \neg \rho_{jk}^{set} \vee \neg guard_k 
\vee \neg \Gamma(r_k) \right )
\end{equation} 

\subsubsection*{\textbf{Formulating cache conflict in set-associative caches}}
With the intuition mentioned in the preceding paragraphs, we formalize the unique 
cache conflict from $r_j$ to $r_i$ via the following logical conditions:  
\begin{equation}
\label{eq:conflict-miss-fifo}
\begin{split}
\Theta_{j,i}^{+,x} \equiv \left ( \phi_{ji}^{cnf,x} \wedge \phi_{ji}^{rel,x} \wedge \phi_{ji}^{eqv,x} 
\wedge guard_j \right ) 
\Rightarrow \left ( \eta_{ji} = 1 \right )
\end{split}
\end{equation} 
\begin{equation}
\label{eq:conflict-hit-fifo}
\begin{split}
\Theta_{j,i}^{-,x} \equiv \left ( \neg \phi_{ji}^{cnf,x} \vee \neg \phi_{ji}^{rel,x} \vee \neg \phi_{ji}^{eqv,x} 
\vee \neg guard_j \right ) 
\\
\Rightarrow \left ( \eta_{ji} = 0 \right )
\end{split}
\end{equation} 
where $x = \{lru,fifo\}$. Concretely, $\eta_{ji}$ is set to $1$ if $r_j$ creates 
a unique cache conflict to $r_i$ and $\eta_{ji}$ is set to $0$ otherwise.

\subsubsection*{\textbf{Computing $\Gamma(r_i)$ for Set-associative Caches}}
To formulate $\Gamma(r_i)$ for set-associative caches, we need to check whether the 
number of unique cache conflicts to $r_i$ exceeds the associativity ($\mathcal{A}$) 
of the cache. 
Based on this intuition, we formalize $\Gamma(r_i)$ for set-associative caches as 
follows: 
\begin{equation}
\label{eq:gamma-set-associative}
\boxed{
\Gamma(r_i) \equiv guard_i \wedge \left ( \Theta_{i}^{cold} \vee \left ( \left ( \sum_{j \in [1,i)} \eta_{ji} \right ) \ge \mathcal{A} \right ) \right ) }
\end{equation}
We note that $\sum_{j \in [1,i)} \eta_{ji}$ accurately counts the number of unique cache 
conflicts 
to the instruction $r_i$ ({\em cf.} \autoref{eq:conflict-miss-fifo}-\autoref{eq:conflict-hit-fifo}). 
Hence, the condition $\left ( \sum_{j \in [1,i)} \eta_{ji} \ge \mathcal{A} \right )$ 
precisely captures whether $\sigma(r_i)$ is replaced from the cache before $r_i$ is executed. 
If $r_i$ does not suffer a cold miss and $\left ( \sum_{j \in [1,i)} \eta_{ji} < \mathcal{A} \right )$, 
then $r_i$ will be a cache hit when executed, as captured by the condition $\neg \Gamma(r_i)$.

\subsection*{Detailed Runtime Monitoring}

\begin{algorithm}[!b!p]
\caption{Monitor extraction and instrumentation}
\label{alg:monitor}
{
\begin{algorithmic}[1]
\Procedure{Monitoring}{$\Psi$, $\objective$, $\varphi$, $\pred$, $\pred_{cur}$, $\Gamma$}
  \State{\textsf{\small /* If $\varphi$ captures side-channel freedom, then $trace$ is}}
  \State{\textsf{\small any of the two traces constituting the counterexample */}}
  \State{$(res, trace)$ := \textsc{Verify}$\left ( \Psi , \varphi \right )$}
  \While{($res$=$false$) $\wedge$ ($trace \neq$ spurious)}
    \State{\textsf{/* Extract observation from $trace$ */}} 
    \State{$o$ := \textsc{GetObservation$\left ( trace \right )$}}
   	\State{ \textsf{/* Extract monitor from $trace$ */}} 
  	\State{$\nu_o$ := $\nu$ := \textsc{ExtractMonitor$\left ( trace \right )$}}
  	\State{ \textsf{/* Refine $\Psi$ to find unique counterexamples */}}
  	\State{$\Psi$ := $\Psi \wedge \neg \nu$}
  	\State{ \textsf{/* Refine $\varphi$ to find all traces exhibiting $o$ */}}
  	\State{$\varphi$ := \textsc{RefineObjective}($\varphi$, $o$)}
  	\State{\textsf{/* Check the unsatisfiability of $\Psi \wedge \neg \varphi$ */}}
  	\State{$(res', trace')$ := \textsc{Verify}$\left ( \Psi , \varphi \right )$}
	\While{($res'$=$false$)}
		\If {($trace' \neq$ spurious)}
  			\State{\label{ln:equivalence-start} $\nu$ := \textsc{ExtractMonitor$\left ( trace' \right )$}}
  			\State{\textsf{/* \small Combine monitors with observation $o$ */}}
  			\State{$\nu_{o}$ := $\nu_{o} \vee \nu$}
  			\State{\textsf{\small /* Refine $\Psi$ for unique counterexamples */}}
  			\State{$\Psi$ := $\Psi \wedge \neg \nu$}
  			\State{\textsf{\small /* Check the unsatisfiability of $\Psi \wedge \neg \varphi$ */}}
  			\State{$(res', trace')$ := \textsc{Verify}$\left ( \Psi , \varphi \right )$}
  		\Else
  			\State{\textsf{\small /* Refine abstraction to repeat verification */}} 
  			\State{\textsc{AbsRefine}$\left ( \Psi, \pred, \pred_{cur}, trace', \Gamma \right )$}
  			\State{\label{ln:equivalence-end} $(res', trace')$ := \textsc{Verify}$\left ( \Psi , \varphi \right )$}
		\EndIf
  	\EndWhile  	
  	\State{Let $\Omega$ holds the set of monitor, observation pairs}
  	\State{$\Omega\ \cup$:= $\{ \langle \nu_o, o \rangle \}$}
  	\State{\textsf{/* Instrument patches for monitor $\nu_o$ */}} 
  	\State{\textsc{InstrumentPatch}($\Omega$, $\objective$)}
  	\State{\textsf{/* Refine objective to find new observations */}} 
  	\State{\label{ln:negate-property} $\varphi$ := $\neg \varphi$}
  	\State{\textsf{ /* Check the unsatisfiability of $\Psi \wedge \neg \varphi$ */}}
  	\State{$(res, trace)$ := \textsc{Verify}$\left ( \Psi , \varphi \right )$}
  \EndWhile
  \State{\textsf{/* Program is still not side-channel free */}} 
  \State{\textsf{/* Refine abstraction to repeat verification loop */}} 
  \If {($res$=$false$)}
  	\State{\textsc{AbsRefine}$\left ( \Psi, \pred, \pred_{cur}, trace, \Gamma \right )$}
  	\State{\textsc{Monitoring$\left (  \Psi, \objective, \varphi, \pred, \pred_{cur}, \Gamma \right )$}}
  \EndIf
 \EndProcedure
 \Procedure{AbsRefine}{$\Psi$, $\pred$, $\pred_{cur}$, $trace$, $\Gamma$}
 	\State{\textsf{/* Extract unsatisfiable core */}} 
  	\State{$\mathcal{U}$ := \textsc{UnsatCore$\left ( trace, \Gamma \right )$}}
  	\State{\textsf{/* Refine abstractions (see \autoref{sec:abstraction}) */}} 
  	\State{$\pred_{cur}$ := \textsc{Refine}($\pred_{cur}$, $\mathcal{U}$, $\pred$)}
  	\State{\textsf{/* Rewrite $\Psi$ with the refined abstraction */}} 
  	\State{\textsc{Rewrite$\left ( \Psi, \pred_{cur} \right )$}}
 \EndProcedure
\end{algorithmic}}
\end{algorithm}
Algorithm~\ref{alg:monitor} outlines the overall process. The procedure \textsc{Monitoring} 
takes the following inputs: 
\begin{itemize}
\item $\Psi$: A symbolic representation of program and cache semantics with the current 
level of abstraction, 

\item $\objective$ and $\varphi$: The model of the attacker ($\objective$) and a property 
$\varphi$ initially capturing cache side-channel freedom w.r.t. $\objective$, 

\item $\pred$ and $\pred_{cur}$: Cache semantics related predicates ($\pred$) and the 
current level of abstraction ($\pred_{cur}$), and 

\item $\Gamma$: Symbolic conditions to determine the cache behaviour.  

\end{itemize}

\subsection*{Additional Experimental Results}

\subsubsection*{\textbf{Sensitivity w.r.t. cache}}
\label{sec:sensitivity}
\autoref{fig:cache-sensitivity} outlines the evaluation for routines that violate side-channel 
freedom and for attack model $\objective_{time}$. 
Nevertheless, the conclusion holds for all routines and attack models. 
\autoref{fig:cache-sensitivity} captures the number of equivalence classes explored 
(hence, the number of patches generated {\em cf.} Algorithm~\ref{alg:monitor}) with 
respect to time. We make the following crucial observations from \autoref{fig:cache-sensitivity}. 
Firstly, the scalability of our checker is stable 
across a variety of cache configurations. This is because we encode cache semantics 
within a program via symbolic constraints on cache conflict. The size of these 
constraints depends on the number of memory-related instructions, but its size 
is not heavily influenced by the size of the cache. Secondly, the number of 
equivalence classes of observations does not vary significantly across cache 
configurations. Indeed, the number of equivalence classes may even increase 
(hence, increased channel capacity) with a bigger cache size 
({\em e.g.} in \texttt{DES} and \texttt{AES}). However, for all cache configurations, 
\CC generated all the patches that need to be applied for making the 
respective programs cache side-channel free. 
\begin{note}
\vspace{-0.2cm}
{\em The scalability of \CC is stable across a variety of cache configurations. 
Moreover, in all cache configurations, \CC 
generated all required patches to ensure the cache side-channel freedom w.r.t. 
$\objective_{time}$.  
}
\vspace{-0.2cm}
\end{note}

\begin{figure*}[!htb]
\begin{center}
\begin{tabular}{ccc}
\rotatebox{0}{
\includegraphics[scale = 0.22]{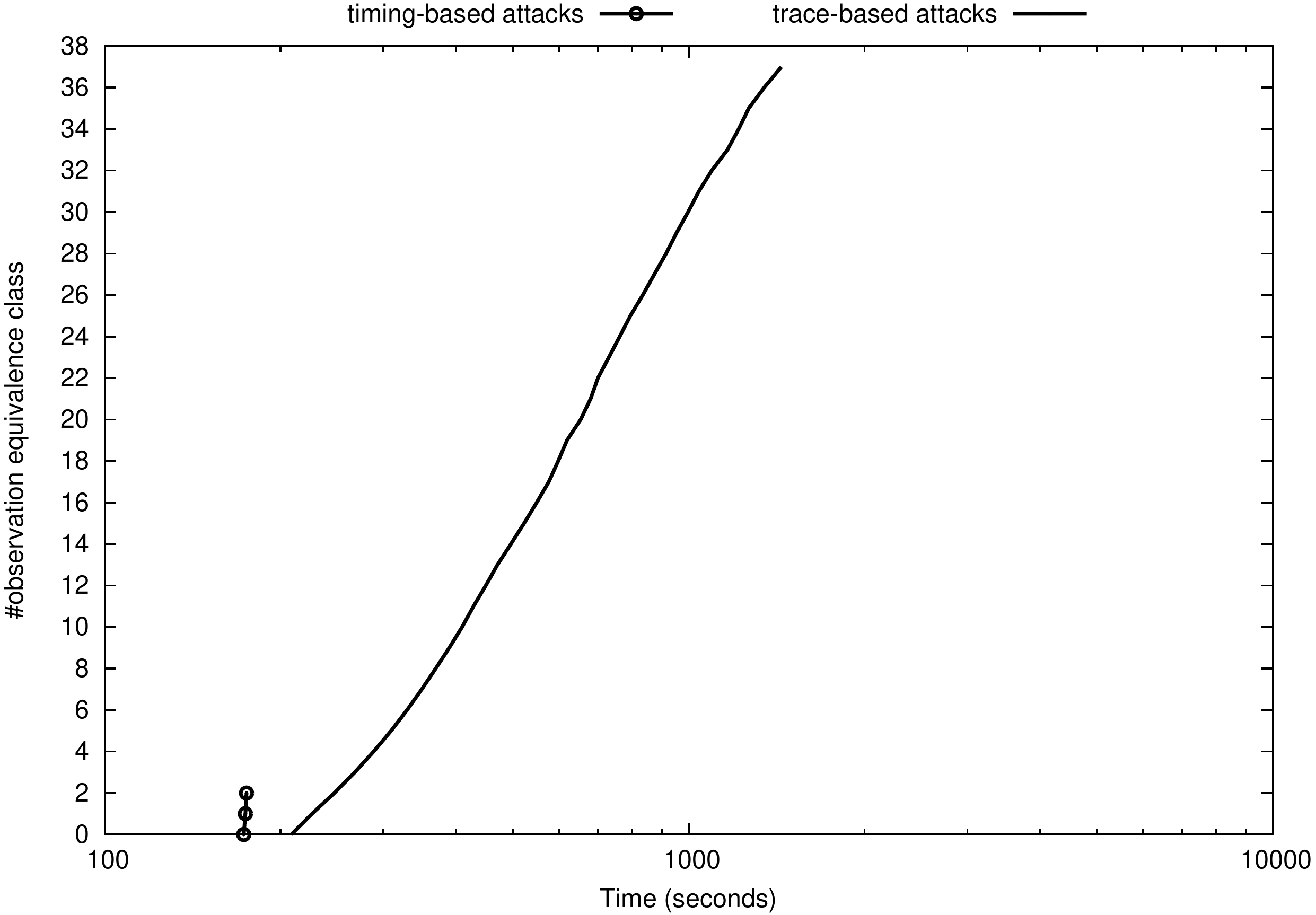}} & 
\rotatebox{0}{
\includegraphics[scale = 0.22]{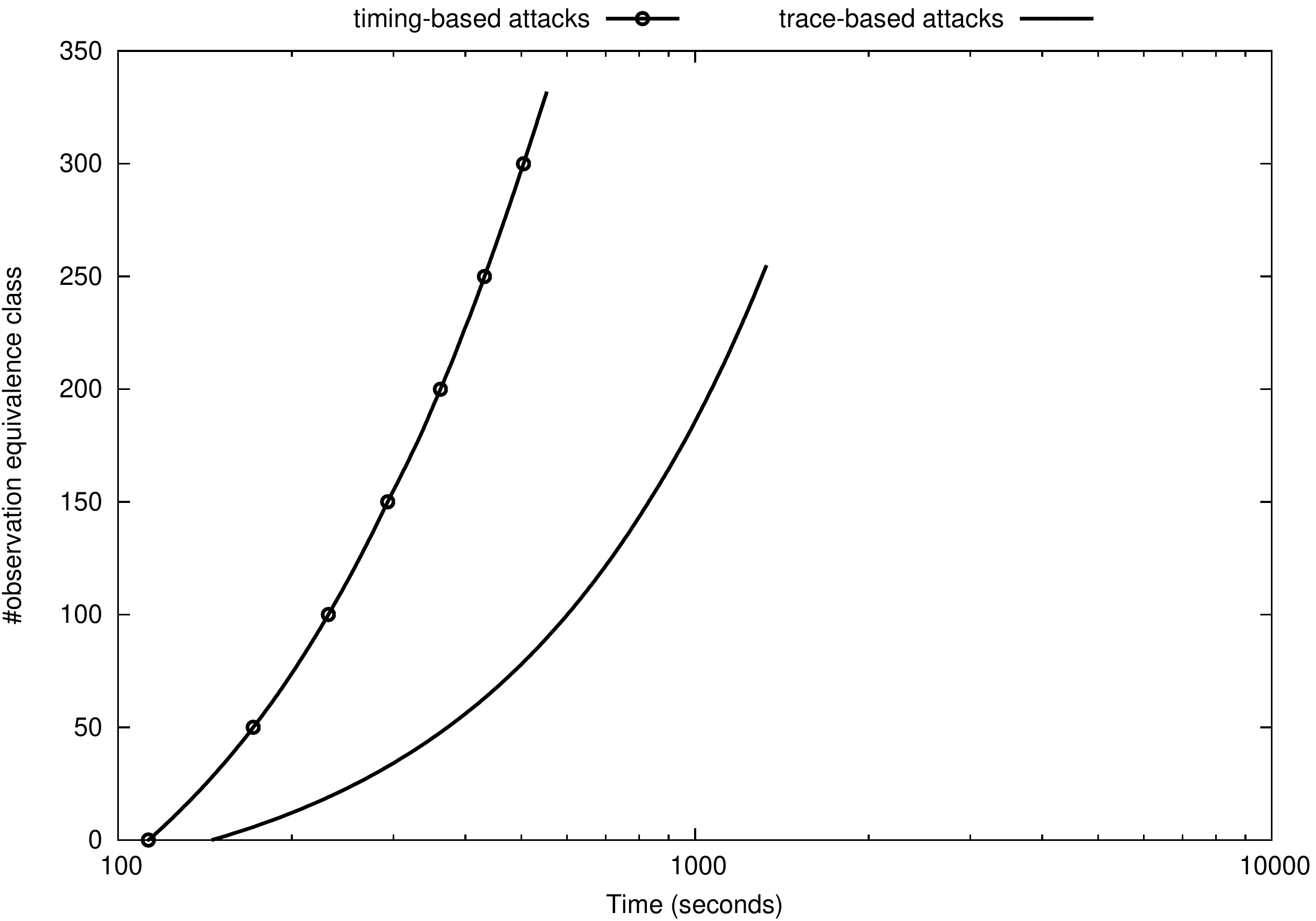}} & 
\rotatebox{0}{
\includegraphics[scale = 0.22]{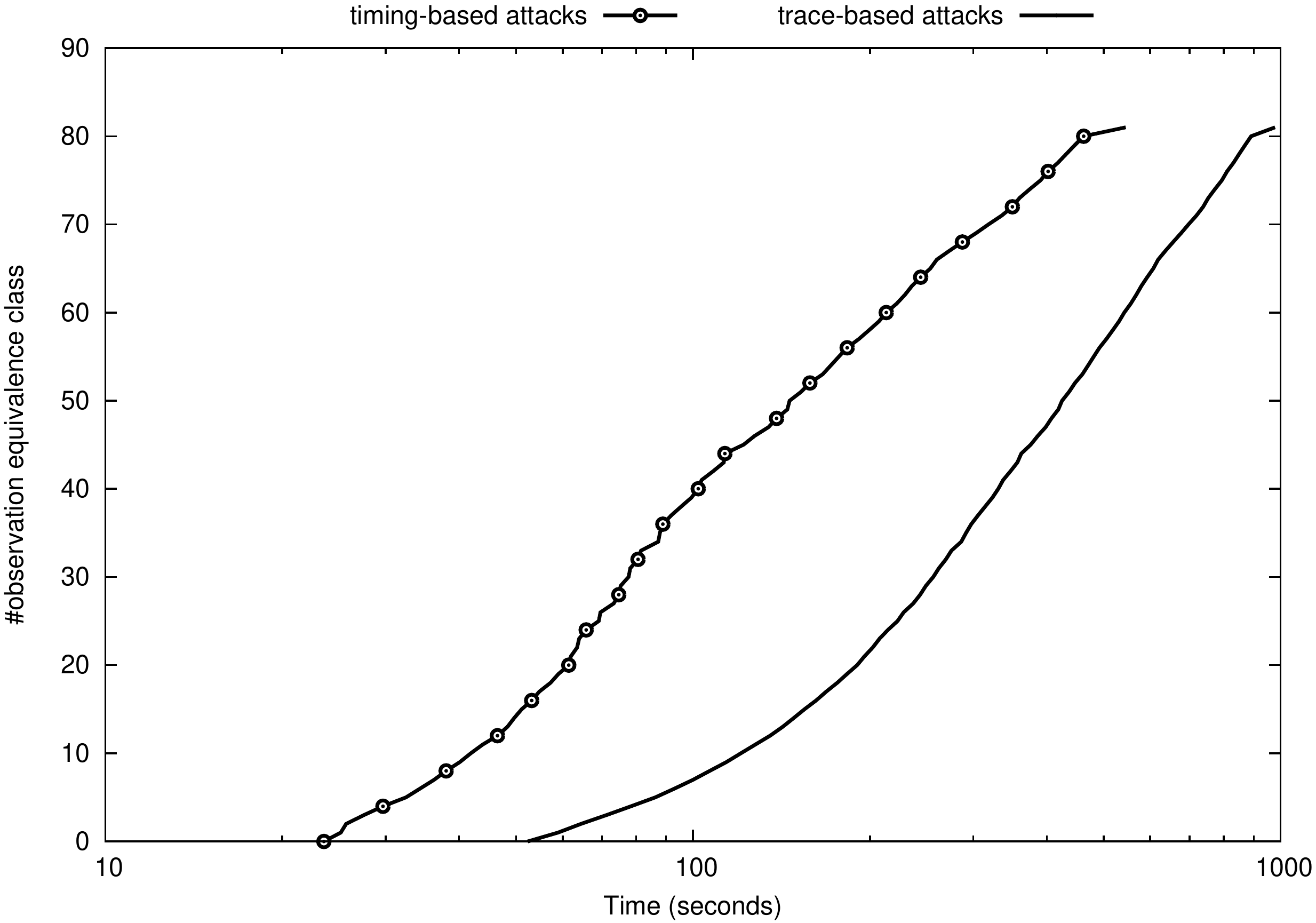}}\\
\textbf{(a)} \texttt{AES128} & \textbf{(b)} \texttt{DES} & \textbf{(c)} \texttt{fix\_frac} \\
\rotatebox{0}{
\includegraphics[scale = 0.22]{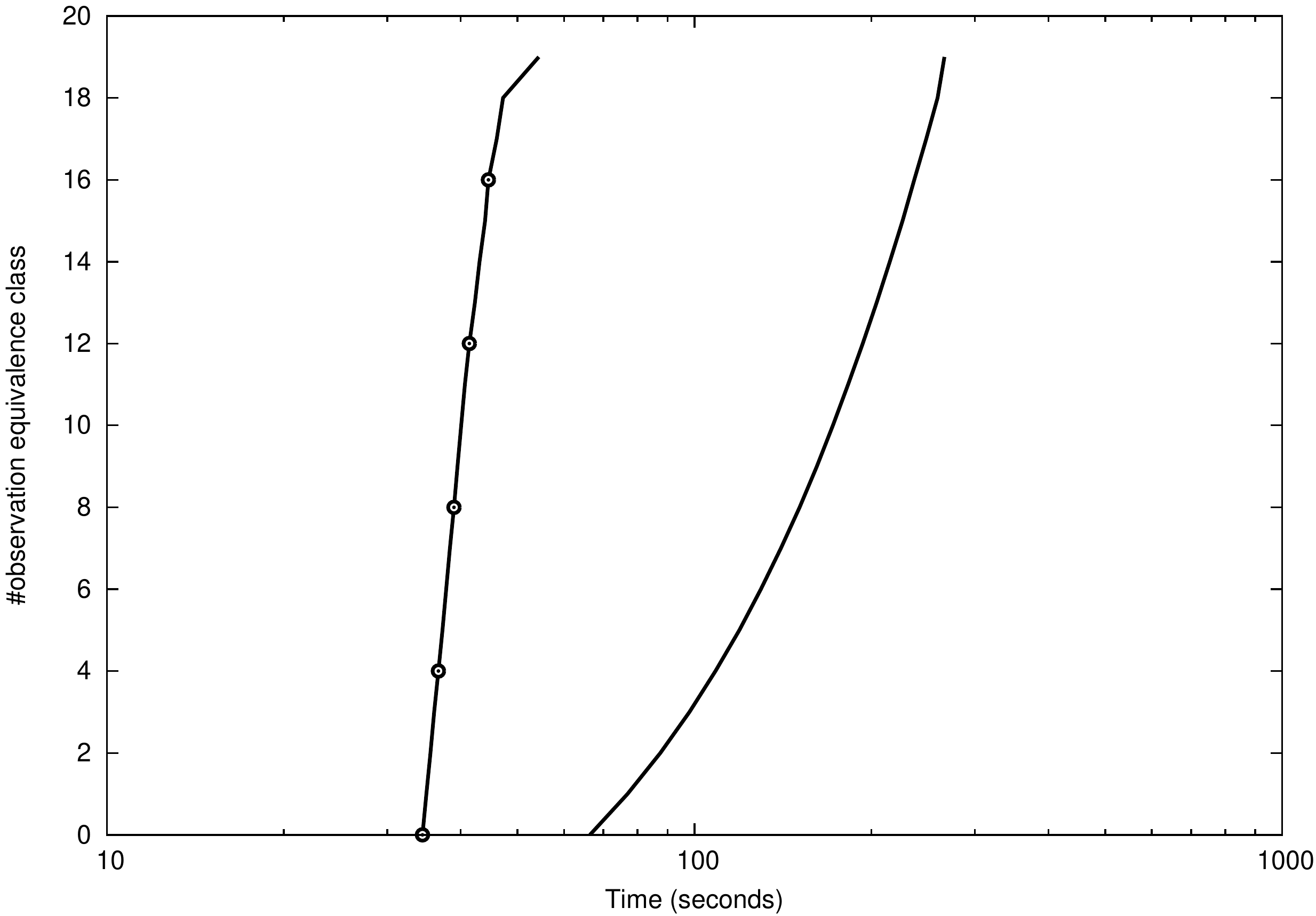}} & 
\rotatebox{0}{
\includegraphics[scale = 0.22]{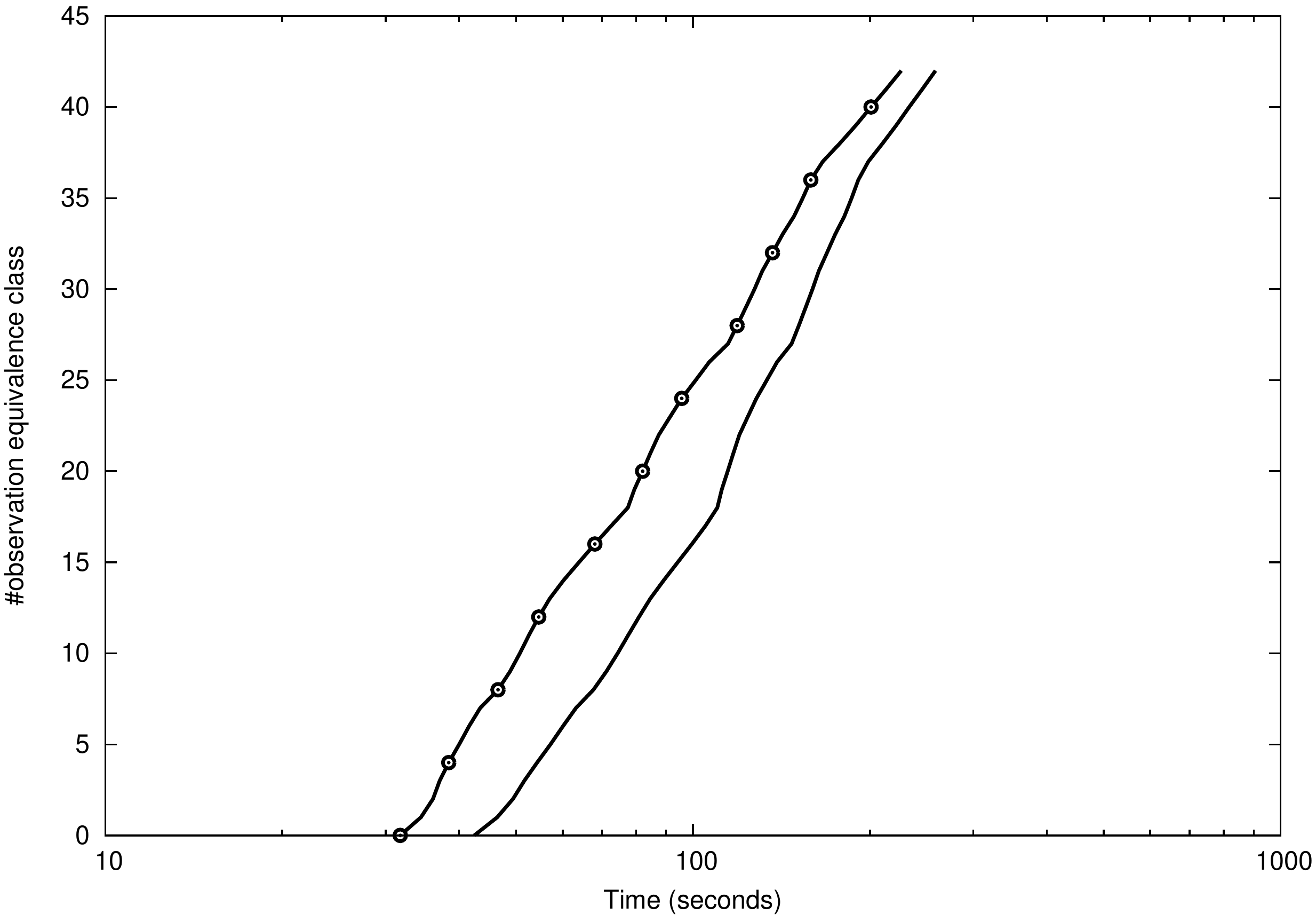}} & 
\rotatebox{0}{
\includegraphics[scale = 0.22]{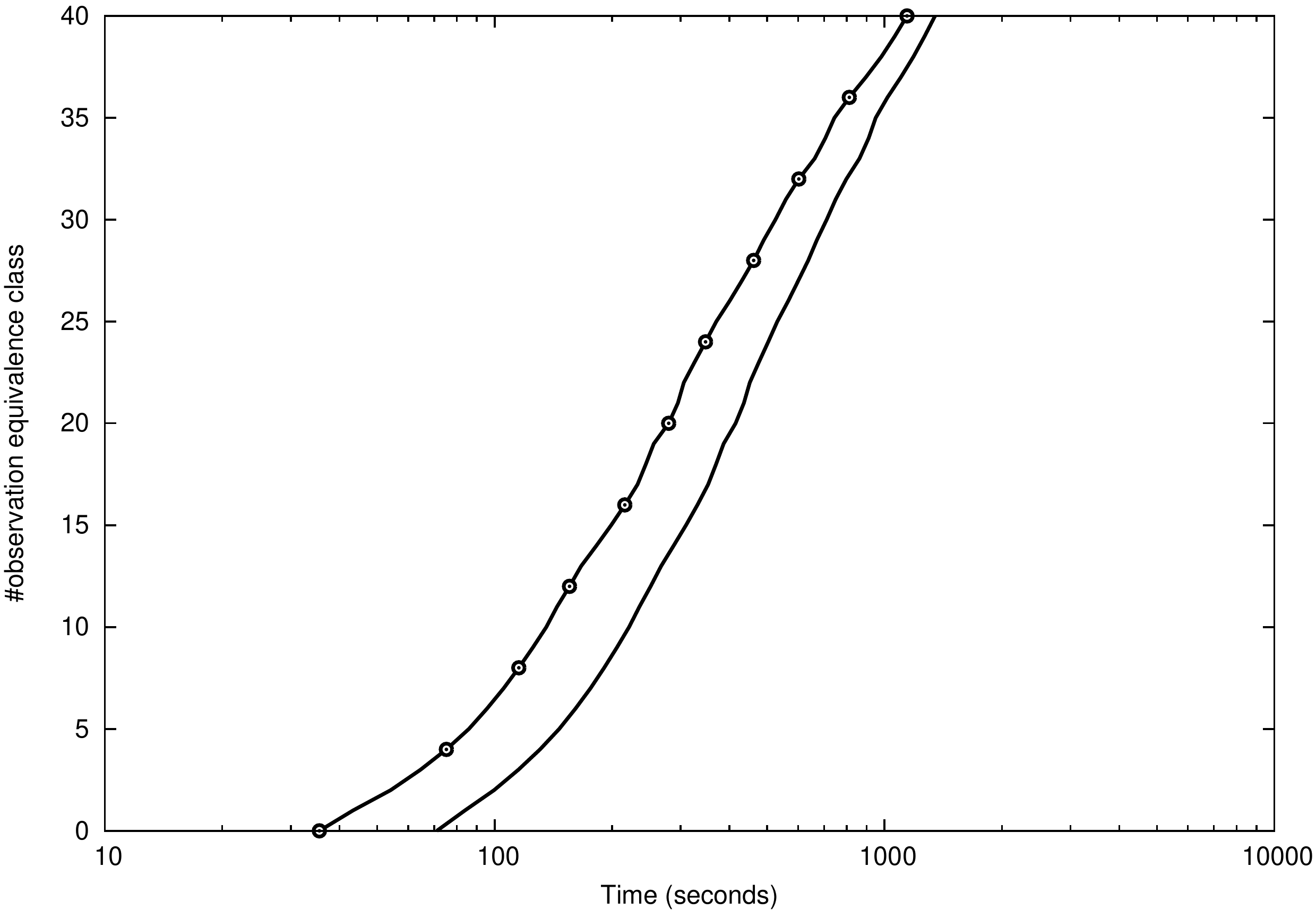}}\\ 
\textbf{(d)} \texttt{eccpoint\_validate} & \textbf{(e)} \texttt{gdk\_unicode\_to\_keyval} & \textbf{(f)} \texttt{gdk\_keyval\_name}\\
\end{tabular}
\end{center}
\vspace*{-0.1in}
\caption{Overhead of counterexample exploration and patch synthesis
}
\label{fig:monitor}
\end{figure*}

\begin{figure*}[!htb]
\begin{center}
\vspace*{-0.1in}
\begin{tabular}{ccc}
\rotatebox{0}{
\includegraphics[scale = 0.22]{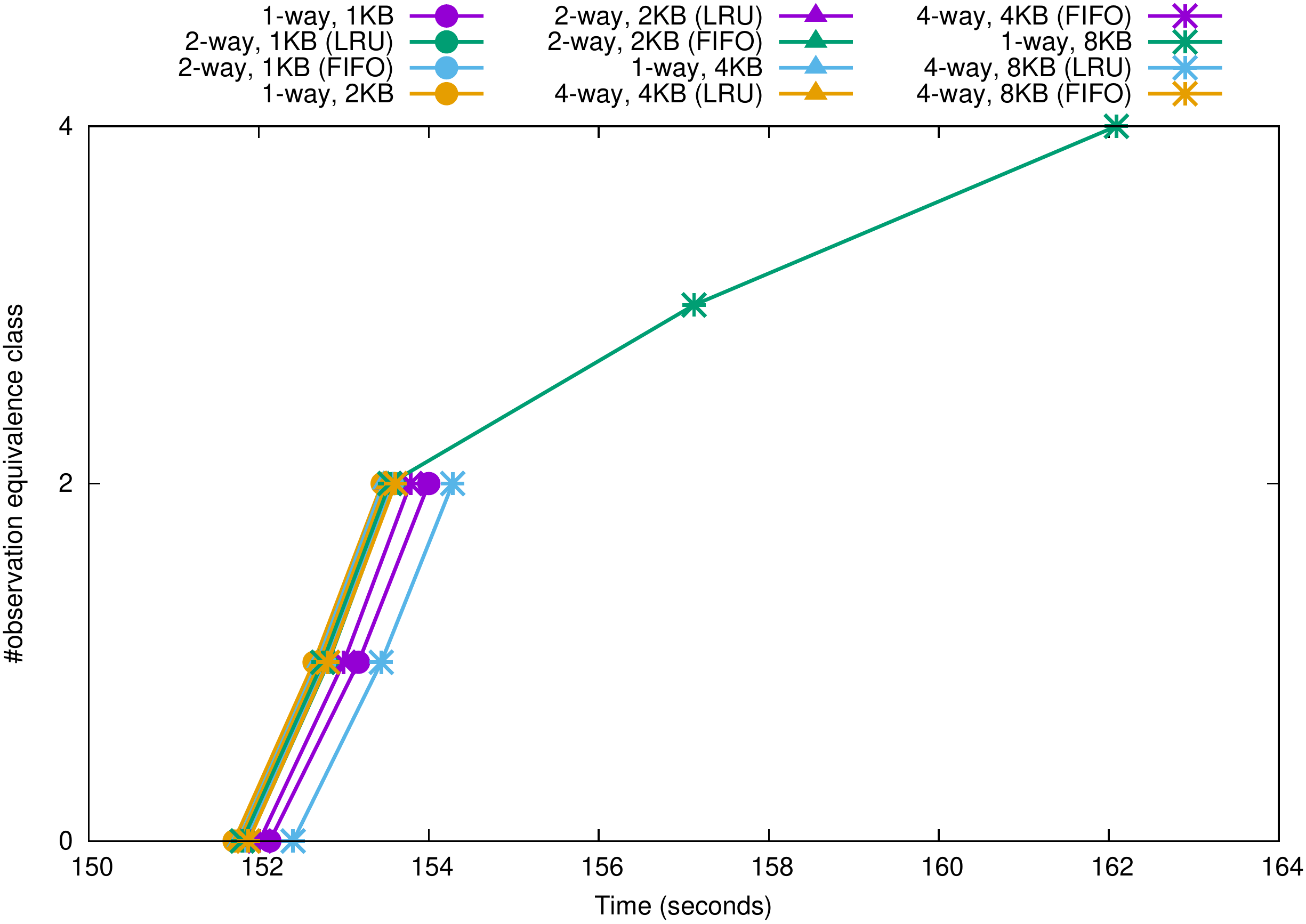}} & 
\rotatebox{0}{
\includegraphics[scale = 0.22]{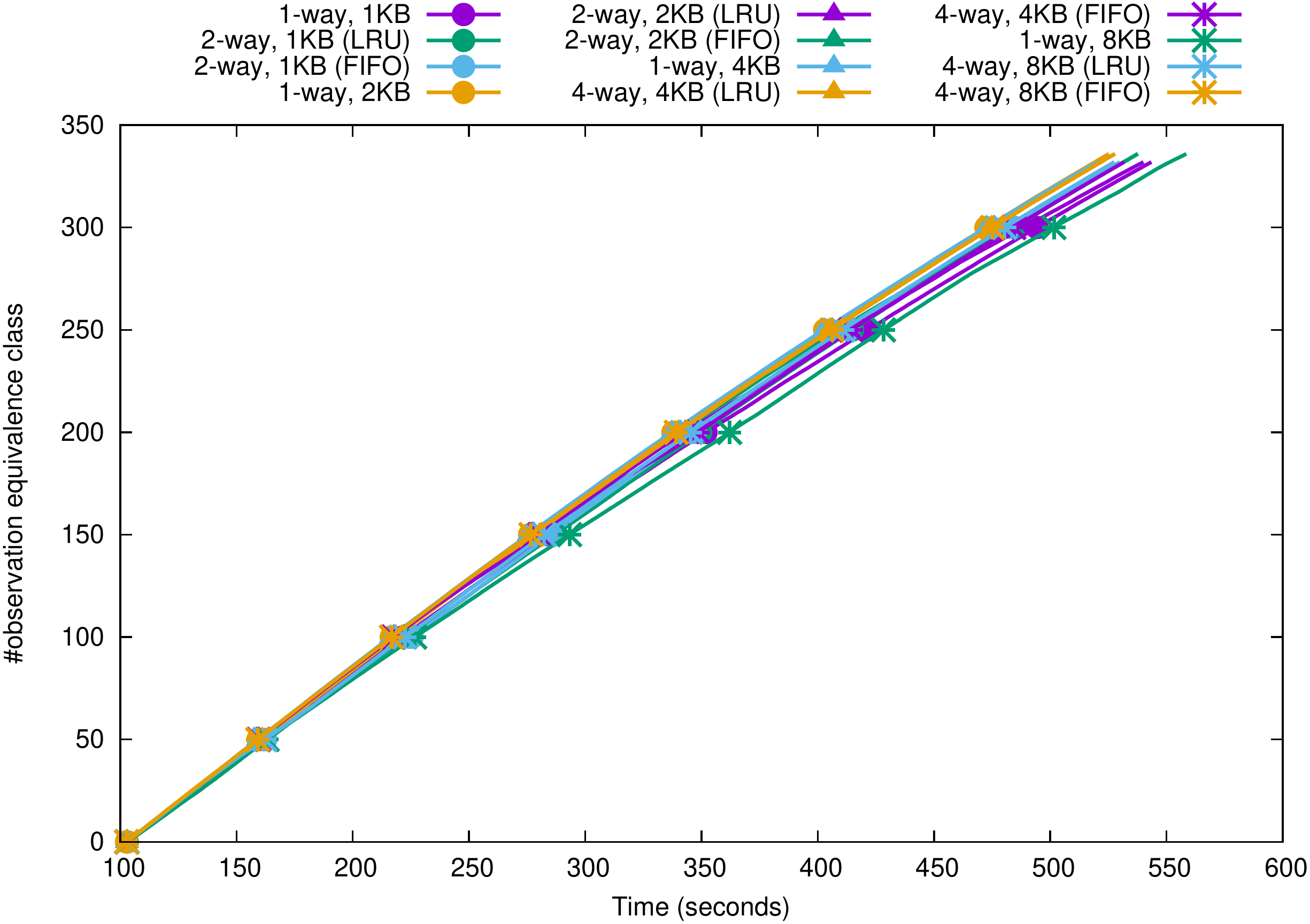}} & 
\rotatebox{0}{
\includegraphics[scale = 0.22]{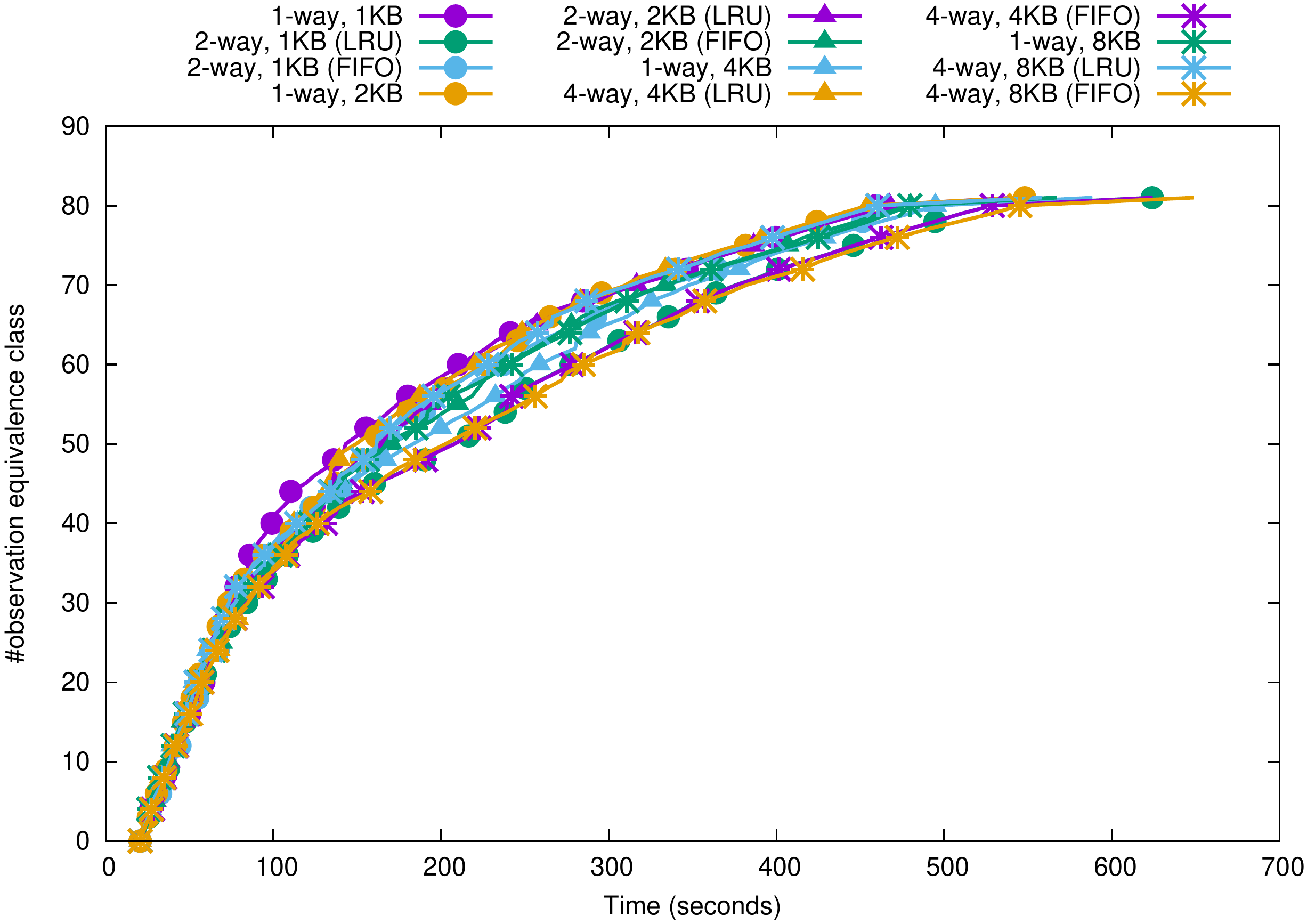}}\\
\textbf{(a)} \texttt{AES128} & \textbf{(b)} \texttt{DES} & \textbf{(c)} \texttt{fix\_frac} \\
\rotatebox{0}{
\includegraphics[scale = 0.22]{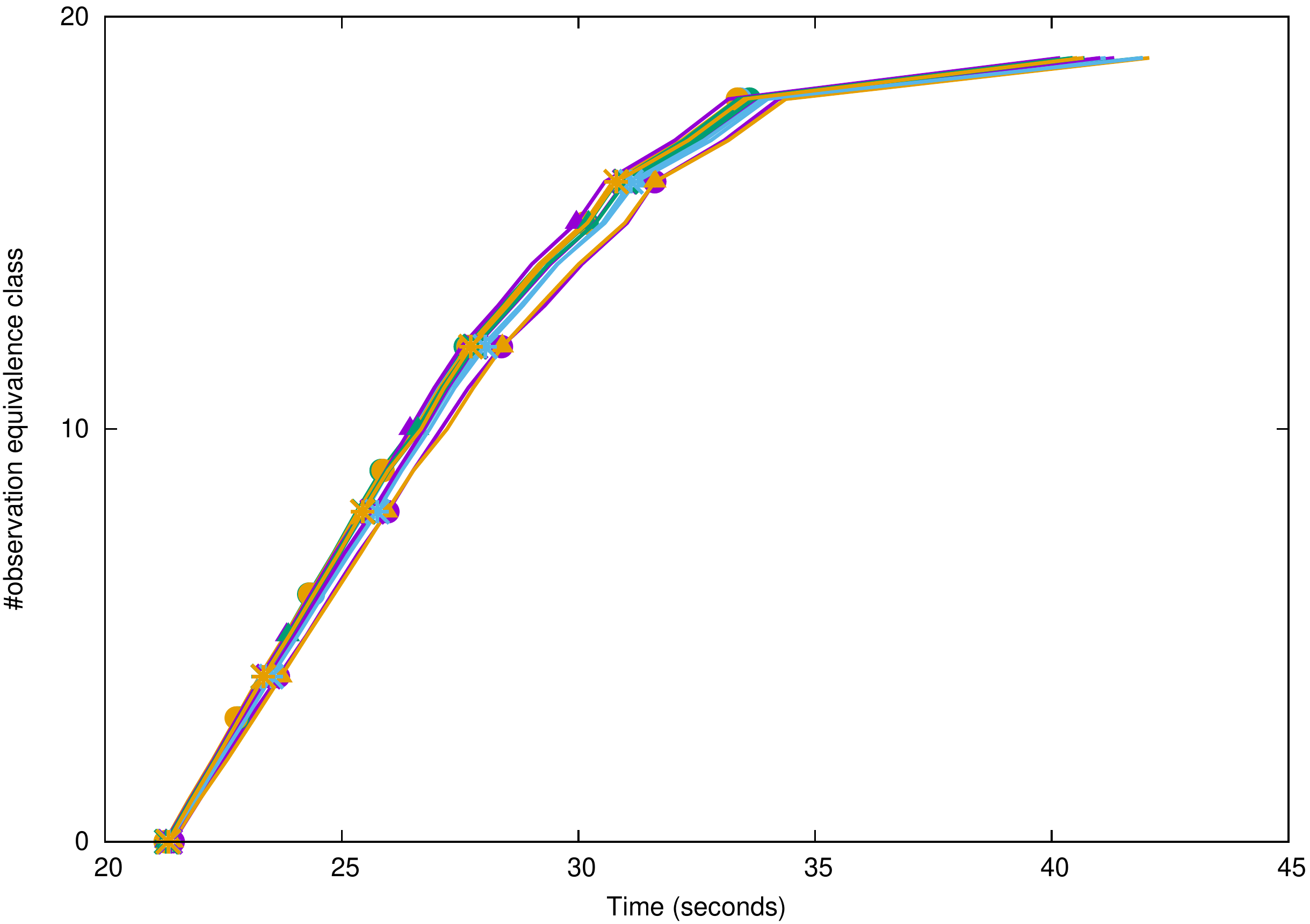}} & 
\rotatebox{0}{
\includegraphics[scale = 0.22]{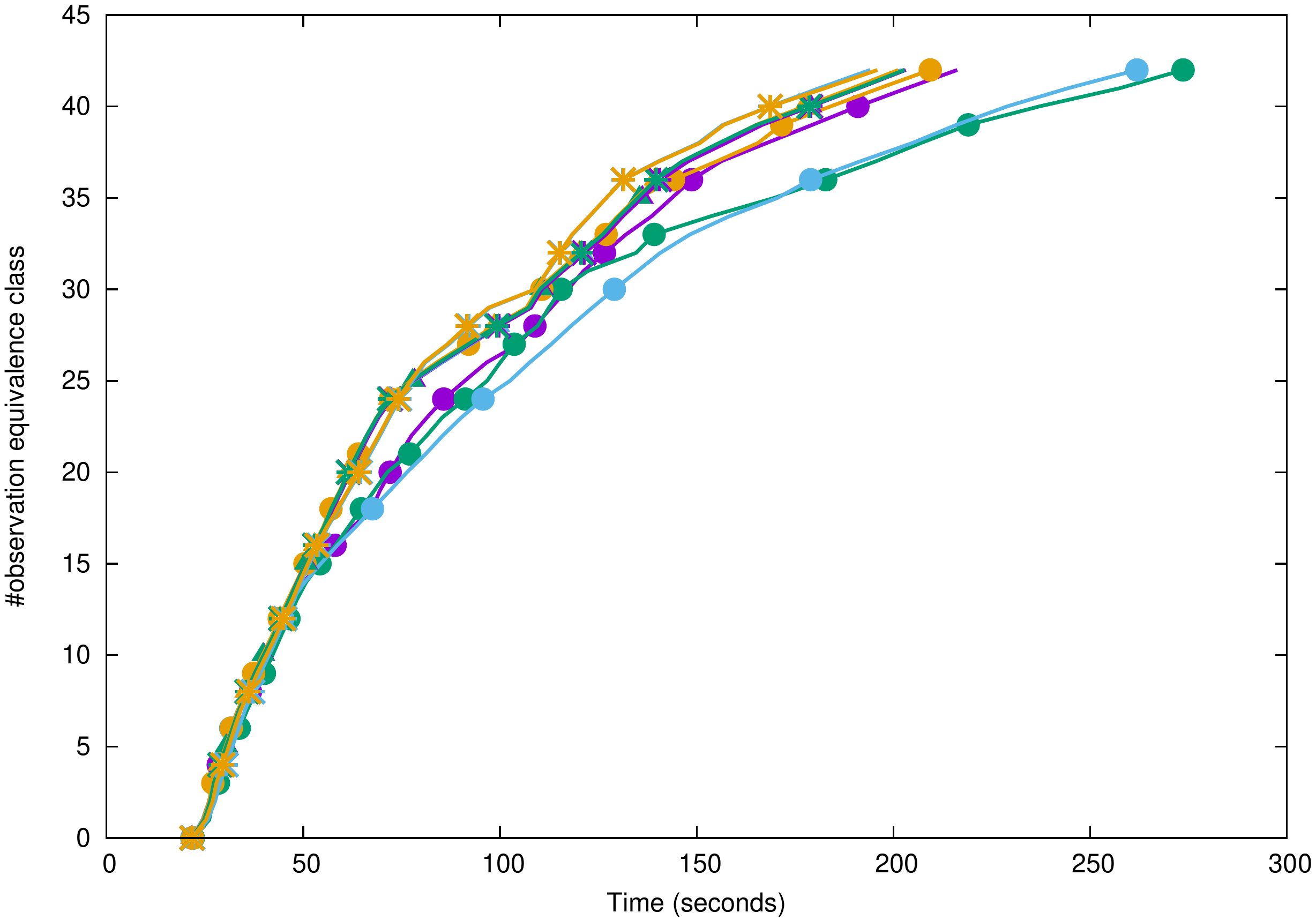}} & 
\rotatebox{0}{
\includegraphics[scale = 0.22]{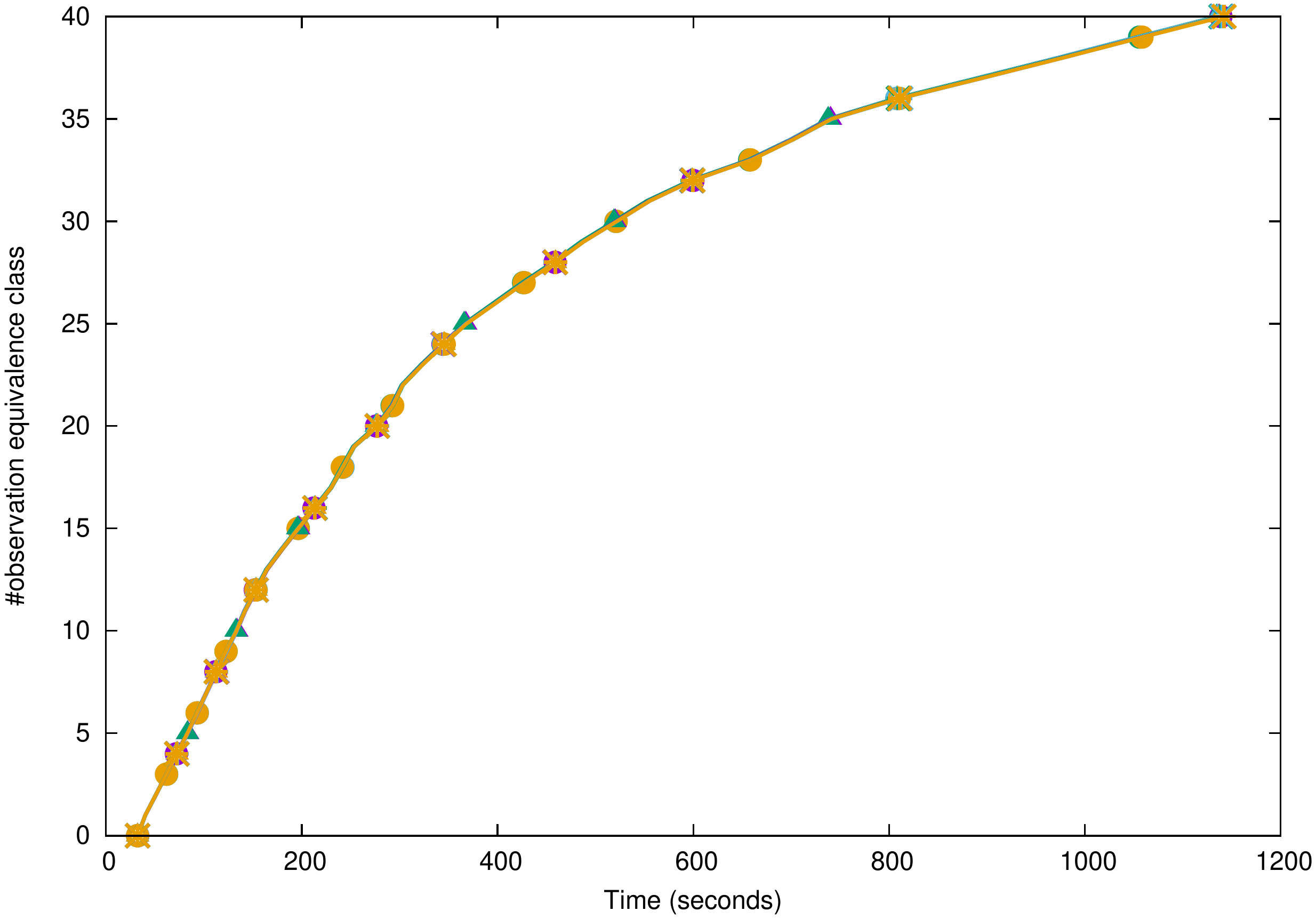}}\\ 
\textbf{(d)} \texttt{eccpoint\_validate} & \textbf{(e)} \texttt{gdk\_unicode\_to\_keyval} & \textbf{(f)} \texttt{gdk\_keyval\_name}\\
\end{tabular}
\end{center}
\vspace*{-0.1in}
\caption{
\CC sensitivity w.r.t. cache 
}
\label{fig:cache-sensitivity}
\vspace*{-0.15in}
\end{figure*}

\end{document}